\documentclass[10pt,journal,compsoc]{IEEEtran}
\pdfoutput=1

\usepackage[utf8]{inputenc}
\usepackage[T1]{fontenc}
\usepackage[style=ieee]{biblatex}
\usepackage{graphicx}
\usepackage[hidelinks]{hyperref}
\usepackage{comment}
\usepackage{listings}
	\lstdefinelanguage{diff}{
    basicstyle=\ttfamily\small,
    morecomment=[f][\color{diffstart}]{@@},
    morecomment=[f][\color{diffincl}]{+\ },
    morecomment=[f][\color{diffrem}]{-\ },
  }
\usepackage[svgnames]{xcolor}
  \definecolor{diffstart}{named}{Grey}
  \definecolor{diffincl}{named}{Green}
  \definecolor{diffrem}{named}{OrangeRed}
\graphicspath{ {./images/} }
\usepackage[inline]{enumitem}
\usepackage{mathtools}
\usepackage{color}
\usepackage{tikz}
\usetikzlibrary{patterns}
\usepackage{pgfplots, pgfplotstable}
\pgfplotsset{compat=1.14}
\usepackage{amsmath}
\usepackage{amssymb}
\usepackage{xspace}
\usepackage{cancel}
\usepackage{beramono}
\usepackage{multirow}
\usepackage{caption}

\definecolor{dkgreen}{rgb}{0,0.6,0}
\definecolor{gray}{rgb}{0.5,0.5,0.5}
\definecolor{mauve}{rgb}{0.58,0,0.82}

\newcommand{\approach}{{\sc SequenceR}\xspace}
\newcommand{\ie}{\textit{i.e.,}\xspace}
\newcommand{\eg}{\textit{e.g.,}\xspace}

\newcommand{\etal}{\textit{et al.}\xspace}
\newcommand{\secref}[1]{Section~\ref{#1}\xspace}
\newcommand{\equref}[1]{Equation~\ref{#1}\xspace}

\newcommand{\figref}[1]{Figure~\ref{#1}\xspace}
\newcommand{\listref}[1]{Listing~\ref{#1}\xspace}

\newcommand{\tabref}[1]{Table~\ref{#1}\xspace}

\bibliography{references.bib}

\makeatletter
\newenvironment{btHighlight}[1][]
{\begingroup\tikzset{bt@Highlight@par/.style={#1}}\begin{lrbox}{\@tempboxa}}
{\end{lrbox}\bt@HL@box[bt@Highlight@par]{\@tempboxa}\endgroup}

\newcommand\btHL[1][]{%
  \begin{btHighlight}[#1]\bgroup\aftergroup\bt@HL@endenv%
}
\def\bt@HL@endenv{%
  \end{btHighlight}%
  \egroup
}
\newcommand{\bt@HL@box}[2][]{%
  \tikz[#1]{%
    \pgfpathrectangle{\pgfpoint{1pt}{0pt}}{\pgfpoint{\wd #2}{\ht #2}}%
    \pgfusepath{use as bounding box}%
    \node[anchor=base west, fill=yellow!30,outer sep=0pt,inner xsep=1pt, inner ysep=0pt, rounded corners=0pt, minimum height=\ht\strutbox+1pt,#1]{\raisebox{1pt}{\strut}\strut\usebox{#2}};
  }%
}
\makeatother
\lstdefinestyle{Highlight}{
    moredelim=**[is][\btHL]{`}{`},
    moredelim=**[is][{\btHL[fill=orange!50]}]{´}{´},
    moredelim=**[is][{\btHL[fill=red!50]}]{@}{@},
}

\begin{document}
\title{\approach: Sequence-to-Sequence Learning for End-to-End Program Repair}

\author{Zimin Chen,
        Steve Kommrusch,
        Michele Tufano, \\ %
        Louis-Noël Pouchet,
        Denys Poshyvanyk
        and
        Martin Monperrus%
\IEEEcompsocitemizethanks{
\IEEEcompsocthanksitem Zimin Chen and Martin Monperrus are with KTH Royal Institute of Technology, 114 28 Stockholm, Sweden\protect\\
E-mail: \{zimin, monp\}@kth.se
\IEEEcompsocthanksitem Steve Kommrusch and Louis-Noël Pouchet are with Colorado State University, Colorado 80523, USA\protect\\
Email: \{steveko, pouchet\}@cs.colostate.edu
\IEEEcompsocthanksitem Michele Tufano and Denys Poshyvanyk are with The College of William and Mary, VA 23185, USA\protect\\
Email: \{mtufano, denys\}@cs.wm.edu
\IEEEcompsocthanksitem Zimin Chen and Steve Kommrusch have equally contributed to the paper as first authors.
}%
\thanks{Manuscript submitted February 11, 2019}}

\markboth{IEEE Transactions on Software Engineering, VOL. TBD, 2019}%
{Chen \etal: \approach: Sequence-to-Sequence Learning for End-to-End Program Repair}
\IEEEtitleabstractindextext{%
\begin{abstract}
This paper presents a novel end-to-end approach to program repair based on sequence-to-sequence learning.
We devise, implement, and evaluate a technique, called \approach, for fixing bugs based on sequence-to-sequence learning on source code. This approach uses the copy mechanism to overcome the unlimited vocabulary problem that occurs with big code. Our system is data-driven; we train it on 35,578 samples, carefully curated from commits to open-source repositories.
We evaluate \approach on 4,711 independent real bug fixes, as well on the Defects4J benchmark used in program repair research.
\approach is able to perfectly predict the fixed line for 950/4,711 testing samples, and find correct patches for 14 bugs in Defects4J benchmark. \approach captures a wide range of repair operators without any domain-specific top-down design.
\end{abstract}

}

\maketitle

\IEEEdisplaynontitleabstractindextext
\IEEEpeerreviewmaketitle

\IEEEraisesectionheading{\section{Introduction}\label{sec:introduction}}
\IEEEPARstart{P}{eople} have long dreamed of machines capable of writing computer programs by themselves.
Having machines writing a full software system is science-fiction but teaching machines to modify an existing program to fix a bug is within the reach of current software technology; this is called automated program repair \cite{Monperrus2015}.

Program repair research is very active and dominated by techniques based on static analysis (\eg Angelix \cite{mechtaev2016angelix}) and dynamic analysis (\eg CapGen \cite{wen2018context}).
While great progress has been achieved, the current state of automated program repair is limited to simple small fixes, mostly one line patches \cite{saha2017elixir,wen2018context}. 
These techniques are heavily top-down, based on intelligent design and domain-specific knowledge about bug fixing in a given language or a specific application domain.
In this paper, we also focus on one line patches, but we aim at doing program repair in a language-agnostic generic manner, fully relying on machine learning to capture syntax and grammar rules and produce well-formed, compilable programs. By taking this approach, we aim to provide a foundation for connecting program repair and machine learning, allowing the program repair community to benefit from training with more complete bug datasets and continued improvements to machine learning algorithms and libraries.

As the foundation for our model, we apply sequence-to-sequence learning \cite{sutskever2014sequence} to the problem of program repair. Sequence-to-sequence learning is a branch of statistical machine learning, mostly used for machine translation: the algorithm learns to translate text from one language (say French) to another language (say Swedish) by generalizing over large amounts of sentence pairs from French to Swedish. The training data comes from the large amount of text already translated by humans, starting with the Rosetta stone written in 196 BC \cite{sole2001rosetta}. 
The name of the technique is explicit: it is about learning to translate from one sequence of words to another sequence of words.

Now let us come back to the problem of programming: we want to learn to 'translate' from one sequence of program tokens (a buggy program) to a different sequence of program tokens (a fixed program). The training data is readily available: we have millions of commits in open-source code repositories.  Yet, we still have major challenges to overcome when it comes to using sequence-to-sequence learning on code:
\begin{enumerate*}
    \item the raw (unfiltered) data is rather noisy; one must deploy significant effort to identify and curate commits that focus on a clear task;
    \item contrary to natural language, misuse of rare words (identifiers, numbers, etc) is often fatal in programming languages \cite{hellendoorn2017deep}; in natural language some errors may be tolerable because of the intelligence of the human reader while in programming languages the compiler (or interpreter) is strict
    \item in natural language, the dependencies are often in the same sentence (``it'' refers to ``dog'' just before) , or within a couple of sentences, while in programming, the dependencies have a longer range: one may use a variable that has been declared dozens of lines before.
\end{enumerate*}

We are now at a tipping point to address those challenges.
First, sequence-to-sequence learning has reached a maturity level, both conceptually and from an implementation point of view, that it can be fed with sequences whose characteristics significantly differ from natural language.
Second, there has been great recent progress on using various types of language models on source code \cite{allamanis2018survey}.
Based on this great body of work, we present our approach to using sequence-to-learning for program repair, which we created to repair real bugs from large open-source projects written in the Java programming language.

Our end-to-end program repair approach is called \approach and it works as follows.
First, we focus on one-line fixes: we predict the fixed version of a buggy programming line. For this, we create a carefully curated training and testing dataset of one-line commits.
Second, we devise a sequence-to-sequence network architecture that is specifically designed to address the two main aforementioned challenges.
To address the unlimited vocabulary problem, we use the copy mechanism \cite{SeeLM17}; this allows \approach to predict the fixed line, even if the fix contains a token that was too rare (\ie an API call that appears only in few cases, or a rare identifier used only in one class) to be considered in the vocabulary. This copy mechanism works even if the fixed line should contain tokens which were not in the training set.
To address the dependency problem, we construct \textit{abstract buggy context} from the buggy class, which captures the most important context around the buggy source code and reduces the complexity of the input sequence. This enables us to capture long range dependencies that are required for the fix.

We evaluate \approach in two ways. First,
we compute accuracy over 4,711 real one-line commits, curated from three open-source projects. The accuracy is measured by the ability of the system to predict the fixed line exactly as originally crafted by the developer, given as input the buggy file and the buggy line number.
Our golden configuration is able to perfectly predict the fix for 950/4,711 (20\%) of the testing samples. This sets up a baseline for future research in the field. 
Second, we apply \approach to the mainstream evaluation benchmark for program repair, Defects4J. Of the 395 total bugs in Defects4J, 75 have one-line replacement repairs; \approach generates patches which pass the test suite for 19 bugs and patches which are semantically equivalent to the human-generated patch for 14 bugs. To our knowledge, this is the first report ever on using sequence-to-sequence learning for end-to-end program repair, including validation with test cases.

Overall, the novelty of this work is as follows. First, we create and share a unique dataset for evaluating learning techniques on one-line program repair. Second, we report on using the copy mechanism on seq-to-seq learning on source code.
Third, on the same buggy input dataset, \approach is able to produce the correct patch for 119\% more samples than the closest related work \cite{tufano2018arxiv}.

To sum up:
\begin{itemize}
    \item Our key contribution is an approach for fixing bugs based on sequence-to-sequence learning on token sequences. This approach uses the copy mechanism to overcome the unlimited vocabulary problem in source code. 
    
    \item We present the construction of an \textit{abstract buggy context} that leverages code context for patch generation. The input program token sequences are at the level of full classes and capture long-range dependencies in the fix to be written. We implement our approach in a publicly-available program repair tool called \approach. 
    
    \item We evaluate our approach on 4,711 real bug fixing tasks. Contrary to the closest related work \cite{tufano2018arxiv}, we do not assume bugs to be in small methods only. Our golden trained model is able to perfectly fix 950/4,711 testing samples. To the best-of-our knowledge, this is the best result reported on such a task at the time of writing this paper \cite{tufano2018arxiv}\cite{Qi2015}\cite{martinez2017automatic}.
    
    \item We evaluate our approach on the 75 one-line bugs of Defects4J, which is the most widely used benchmark for evaluating programming repair contributions.
    \approach is able to find 2,321 patches for these bugs, 761 compile successfully, 61 are plausible (they pass the full test suite) and 18 are semantically equivalent to the patch written by the human developer.
    
    \item We provide a qualitative analysis of 8 interesting repair operators captured by sequence-to-sequence learning on the considered training dataset.

\end{itemize}

\section{Background on neural machine translation with sequence-to-sequence learning}
\label{sec:nmt}
\approach is based on the idea of receiving buggy code as input and producing fixed code as output. The concept is similar to neural machine translation where the input is a sequence of words in one language and the output is a sequence in another language. In this section, we provide a brief introduction to neural machine translation (NMT).

\begin{figure}
    \centering
    \includegraphics[width=0.5\textwidth]{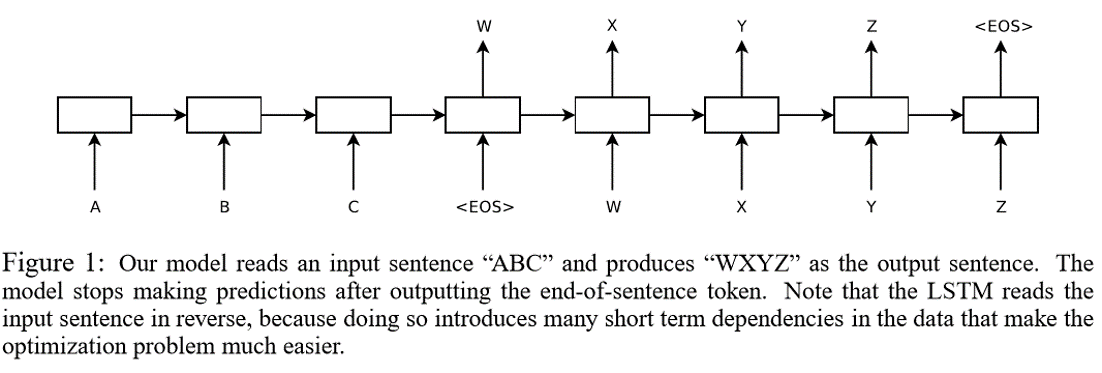}
    \caption{Figure from Sutskever, \etal \cite{sutskever2014sequence} showing example of early sequence-to-sequence model}
    \label{fig:initseq2seq}
\end{figure}

In neural machine translation, the dominant technique is called ``sequence-to-sequence learning'', where ``sequence'' refers to the sequence of words in a sentence.
An early example of a sequence-to-sequence network \cite{sutskever2014sequence} used a recurrent neural network to read in tokens and to generate an output sequence, as shown in \figref{fig:initseq2seq}. 
Let us consider that the input tokens are denoted $x_t$, and after receiving all of the input tokens a special <EOS> token is used. The output tokens are denoted $y_t$, and at training time the output tokens are fed into the network to learn proper generation of the next token. 
In the following equations, $h_t$ is the hidden state of a recurrent neural network, $W^{hx}$ is the weight matrix that computes how the input $x_t$ affects the hidden state, $W^{hh}$ is the weight matrix related to recurrence (\ie how the previous hidden state affects the current hidden state), and $W^{yh}$ is the weight matrix used to predict which token should be output given the hidden state. All weights are learned with supervised learning and back-propagation:
\begin{center}
$h_t = \sigma (W^{hx}x_t + W^{hh}h_{t-1}) $ \\
$y_t = W^{yh}h_t  $
\end{center}

A softmax function is then used to turn the $y_t$ values into probabilities to choose the most likely token from a learned vocabulary. In this example, one can see how the weight matrices capture the learning of common patterns; after processing the input sequence, the hidden state $h_{<eos>}$ encodes the most likely initial token to begin the output and each subsequent $h_t$ uses the $W$ matrices to predict the most likely next token given the input as well as preceding tokens just produced in the output. The $W$ matrices thus learn the long range dependencies in the full input.

A problem with the sequence generation described above is that only tokens which are in the training set are  available for output as $y_t$. In the case of natural human language, words such as proper names (\eg Chicago, Stockholm) may be so rare that they do not appear in the training vocabulary, but those words may be necessary for proper output. One successful approach to overcome the vocabulary problem is to use a copy mechanism \cite{SeeLM17}. The basic intuition behind this approach is that rare words not available in the vocabulary (\ie unknown words, referred as \texttt{<unk>}), may be directly copied from the input sentence over to the output translated sentence. This relatively simple idea can be successful in many cases - especially when translating sentences containing proper names - where these tokens can be easily copied over.

For example, let's consider the task of translating the following English sentence \textit{"The car is in Chicago"} to French. Let's also assume that all the tokens in the sentence are in the vocabulary, except \textit{"Chicago"}. An NMT model might output the following sentence: \textit{"La voiture est \`a \texttt{<unk>}"}. With a copy mechanism, the model would be able to automatically replace the unknown token with one of the tokens from the input sentence, in this case, \textit{"Chicago"}.

The copy mechanism can be particularly relevant for source code, where the size of the vocabulary can be several times the size of a natural language corpus \cite{Tufano-ICSE19}. This results from the fact that developers are not constrained by any vocabulary (\eg English dictionary) when defining names for variables or methods. This leads to an extremely large vocabulary containing many rare tokens, used infrequently only in specific contexts. Thus, the copy mechanism applied to source code allows a system to generate rare out-of-vocabulary identifier names and numeric values as long as they are somewhere in the input. Furthermore, in natural language, a human recipient may be able to use context to cope with one missing word in an automatically translated sentence. In a programming language, the compiler does not make any semantic inference, and the generation has to be complete. For example, if the code to predict is "if (i < num\_cars)", then generating "if (i < int)" is not going to work at all.
We discuss the mathematics of the copy mechanism in the context of \approach in \secref{sec:model}. Readers interested in more detail are referred to the work by See et al. \cite{SeeLM17}.

Tufano \etal \cite{tufano2018arxiv} proposed using NMT with the goal of learning bug-fixing patches by translating the entire buggy method into the corresponding fixed method. Before the translation, the authors perform a code abstraction process which transforms the source code into an abstracted version, which contains: (i) Java keywords and identifiers; (ii) frequent identifiers and literals (a selection of 300 idioms); (iii) typified IDs (\eg \texttt{METHOD\_1}, \texttt{VAR\_2}) that replace identifiers and literals in the code. In \secref{sec:related} we highlight differences and improvements introduced in \approach.

Another approach to addressing the vocabulary size problem in code is to use byte pair encoding (BPE), which has been widely used in NLP and also applied to source code \cite{Karampatsis19}. For \approach, we did preliminary experiments with BPE to solve the unlimited vocabulary problem, but our early results showed that it is less effective than the copy mechanism.

\begin{figure*}[h]
    \centering
    \includegraphics[width=0.9\textwidth]{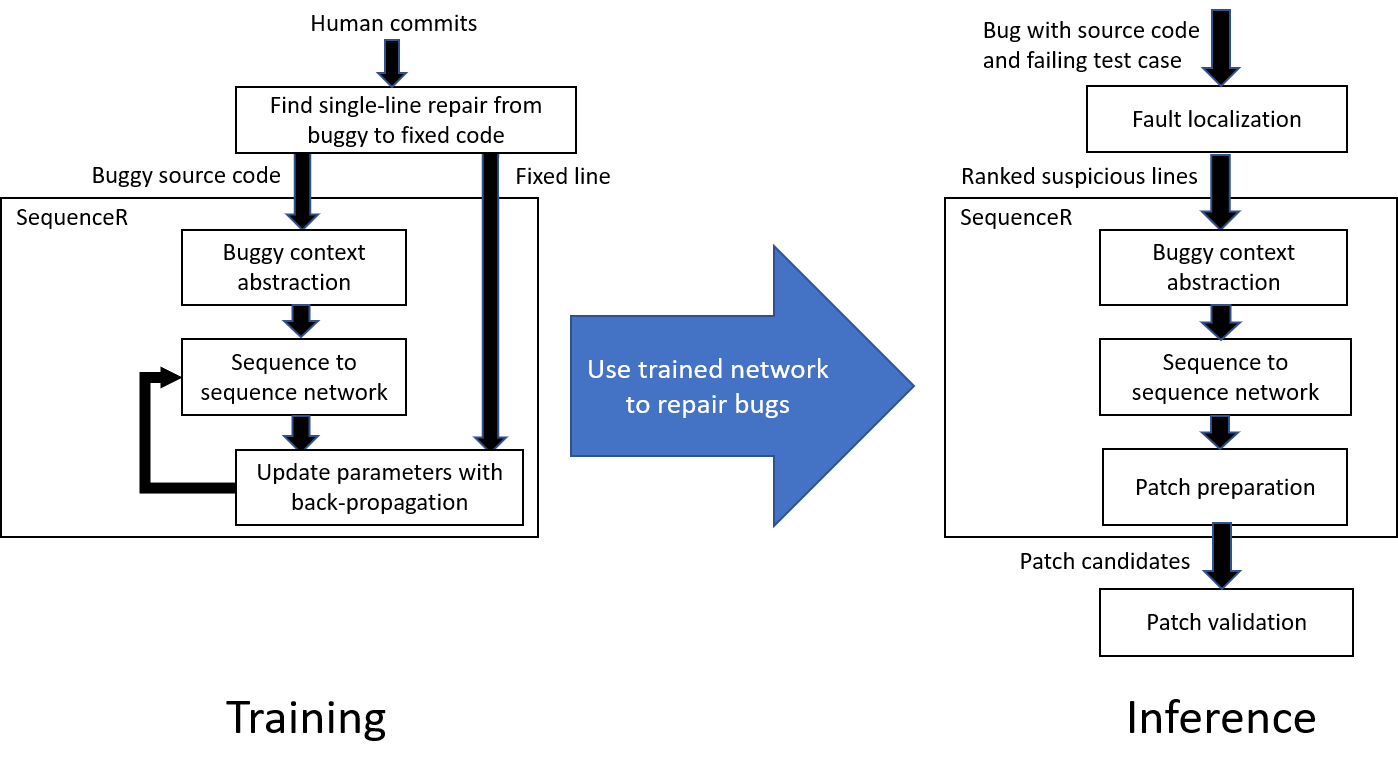}
    \caption{Overview of our approach using sequence-to-sequence learning for program repair.}
    \label{fig:end2end}
\end{figure*}

\begin{figure*}[ht]
\begin{center}
\begin{minipage}[t]{.3\textwidth}
\begin{lstlisting}[language=Java,columns=flexible, frame=single, basicstyle=\small,label={lst:fooOriginal},caption={Original code},captionpos=b,breaklines=true,tabsize=2,numbers=left,xleftmargin=3em,framexleftmargin=2em,keywordstyle=\color{blue},commentstyle=\color{dkgreen},stringstyle=\color{mauve},literate={\ \ }{{\ }}1,style=Highlight]
`class Foo {`
`   int i = 0;`
`   int bar;`
`   Foo (int bar){`
`       this.bar = bar;`
`   }`
`   int decrement(){`
`       return bar-1;`
`   }`
´   int increment(){´
@       return bar-1;@
´   }´
`}`
\end{lstlisting}
\end{minipage}
\hfill
\begin{minipage}[t]{.3\textwidth}
\begin{lstlisting}[language=Java,columns=flexible, frame=single, basicstyle=\small,label={lst:buggyContext},caption={\textit{abstract buggy context}},captionpos=b,breaklines=true,tabsize=2,numbers=left,xleftmargin=3em,framexleftmargin=2em,keywordstyle=\color{blue},commentstyle=\color{dkgreen},stringstyle=\color{mauve},literate={\ \ }{{\ }}1,style=Highlight]
`class Foo {`
`   int i = 0;`
`   int bar;`
`   Foo (int bar){`
`   }`
`   int decrement(){`
`   }`
´   int increment(){´
@       <START_BUG> @
@       return bar-1; @
@       <END_BUG>@
´   }´
`}`
\end{lstlisting}
\end{minipage}
\hfill
\begin{minipage}[t]{.3\textwidth}
\begin{lstlisting}[language=Java,columns=flexible, frame=single, basicstyle=\small,label={lst:fooAbstracted},caption={Context with <unk>},captionpos=b,breaklines=true,tabsize=2,numbers=left,xleftmargin=3em,framexleftmargin=2em,keywordstyle=\color{blue},commentstyle=\color{dkgreen},stringstyle=\color{mauve},literate={\ \ }{{\ }}1,style=Highlight]
`class <unk> {`
`   int i = 0;`
`   int <unk>;`
`   <unk> (int <unk>){`
`   }`
`   int <unk>(){`
`   }`
´   int increment(){´
@       <START_BUG> @
@       return <unk>-1; @
@       <END_BUG>@
´   }´
`}`
\end{lstlisting}
\end{minipage}
\end{center}
\caption{Illustration of the \textit{abstract buggy context} step in \approach. $b^c$ is highlighted in yellow, $b^m$ is highlighted in orange and $b^l$ is highlighted in red.}
\label{fig:abstraction}
\end{figure*}

\section{Approach to Using Seq-to-Seq Learning for Repair}
\approach is a sequence-to-sequence deep learning model that aims at automatically fixing bugs by generating one-line patches (\ie the bug can be fixed by replacing a single buggy line with a single fixed line).
We do not consider line deletion because: 1) it does not require a method for token generation (and is thus less interesting to our research) and 2) if desired, \approach could be combined with the lightweight Kali \cite{Qi2015} to include line deletion. We do not consider line addition because spectrum based fault localization, used in most of the related work, is not effective for line addition patches \cite{ZouLXEZ2019}. We note that in 64\% of all 395 bugs in Defects4J are fixed by replacing existing source code \cite{just2014defects4j}.
Given a Software System with a faulty behavior (\ie failing test case), state-of-the-art fault localization techniques are used to identify the buggy method and the suspicious buggy lines. Such techniques have been shown to predict the correct buggy line as one of the top 10 candidates in 44\% of the time \cite{ZouLXEZ2019}. \approach then performs a novel \textit{\textbf{Buggy Context Abstraction}} (\secref{sec:context-abstraction}) process which intelligently organizes the fault localization data (\ie buggy classes, methods, and lines) into a representation that is concise and suitable for the deep learning model yet able to preserve valuable information regarding the context of the bug, which will be used to predict the fix. The representation is then fed to a trained sequence-to-sequence model (\secref{sec:model}) which performs \textit{\textbf{Patch Inference}} (\secref{sec:patch-inference}) and is capable of generating multiple single-lines of code that represent the potential one-line patches for the bug. Finally, \approach in the \textit{\textbf{Patch Preparation}} (\secref{sec:patch-preparation}) step generates the concrete patches by formatting the code and replacing the suspicious line with the proposed lines. \figref{fig:end2end} shows the aforementioned steps both for the training phase (left) and inference phase (right). In the remainder of this section we will discuss the common steps as well as those specific for training and inference.

\subsection{Problem Definition}
Given a buggy system $b^s$, and test suite $t$, we assume a fault localization technique, $FL$, which identifies an ordered set of potential bug locations $l = \{l_1, l_2,...\}$, where each location $l_i$ consists of the buggy class $b_i^c$, buggy method $b_i^m$, and the buggy line $b_i^l$:
\begin{align*}
l &= \{loc \mid loc \in FL(b^s, t)\} \\
\forall l_i \in l, l_i &= \{b_i^c, b_i^m,b_i^l\} \quad \text{and} \quad b_i^l \subset b_i^m \subset b_i^c
\end{align*}
The problem is to predict (\ie generate) a fixed line $f_i^l$, where $l_i$ is the true bug location, such that by replacing $b_i^l$ with $f_i^l$ in $b_i^m$, the resulting system $f^s$ passes the test suite and the bug is considered fixed.
\approach tackles this problem by taking as input the fault localization data (\ie $l = \{l_1, l_2,...\}$) of a buggy system and attempts to generate fixed line $f_i^l$ for each $l_i$ in order. The $b^s$, $t$, $l$, $l_i$, $b_i^c$, $b_i^m$, $b_i^l$, $f_i^l$ and $f^s$ notations are used throughout this work.

\subsection{Buggy Context Abstraction}
\label{sec:context-abstraction}
The \textit{context} of a bug plays a fundamental role in understanding the faulty behavior and reasoning about the possible fix. During bug-fixing activities, developers usually identify the buggy lines, then analyze how they interact with the rest of the method's execution, and observe the context (\eg variables and other methods) in order to reason about the possible fix and possibly select several tokens in the context to build the fixed line \cite{ko2006exploratory}. 

\approach mimics this process by constructing the \textit{abstract buggy context} and organizing the fault localization data into a representation that is concise yet retains the necessary context that allows the model to predict the possible fix. During this process \approach needs to balance two contrasting goals: (i) reduce the buggy context into a reasonably concise sequence of tokens (since sequence-to-sequence models suffer from long sentences \cite{DBLP:journals/corr/ChoMBB14}), (ii) while at the same time retaining as much information as possible to allow the model to have enough context to predict a possible fix.

Given the bug locations $l = \{l_1,l_2,...\}$, for each $l_i \in l, l_i = \{b_i^c, b_i^m, b_i^l\}$, \approach performs the following steps:

\begin{description}
  \item[Buggy Line]
  \texttt{<START\_BUG>} is inserted before the first token in the buggy line $b_i^l$ and \texttt{<END\_BUG>} is inserted after the last token.
  The rationale is that we would like to propagate the information extracted by the fault localization technique and indicate to the model what is a buggy line. In doing so, we mimic  developers who focus on the buggy lines during their bug-fixing activities.
  \item[Buggy Method] The remainder of the buggy method $b_i^m$ is kept in the representation. The rationale is that the method provides crucial information on where the buggy line is placed and its interaction with the rest of the method.
  \item[Buggy Class] 
  From the buggy class $b_i^c$ we keep all the instance variables and initializers, along with the signature of the constructor and non-buggy methods even if they are not called in the buggy method. The body of the non-suspcious methods is stripped out.
  The rationale for this choice is that the model could use variables and method signatures as potential sources when building the fixed line $f_i^l$.
\end{description}

After these steps, \approach performs tokenization and truncation to create the \textit{abstract buggy context}. 
Truncation is used to limit the \textit{abstract buggy context} to a predetermined size in cases where the input sequence is too long. This allows \approach to process input files of arbitrary size without running out of memory. The truncation process can be summarized as: 
\begin{enumerate*}
\item the truncation size will be chosen such that most input files do not require truncation
\item if the buggy line itself is over the truncation limit, as many tokens as possible from the start of the line are included up to the limit
\item otherwise, the buggy line is included in \textit{abstract buggy context} and twice as many tokens are included before the line as after the line.
\end{enumerate*} For example, if the truncation limit is 1,000 tokens and a 5,000 token file has a buggy line with 100 tokens (including the START\_BUG and END\_BUG tokens) in the middle of the file, then \textit{abstract buggy context} will consist of 600 tokens before the buggy line, then 100 tokens of the buggy line, then 300 tokens after the buggy line. Generally, truncation will delete the actual class definition from the input, but context near the buggy line is preserved to aid in patch generation.

The \textit{abstract buggy context} represents the input to the sequence-to-sequence network which will be used to predict the fixed line. Internally, \textit{abstract buggy context} is represented as a sequence of tokens belonging to a vocabulary $V$. The out-of-vocabulary tokens ($token \not\in V$) are replaced with the unknown token \texttt{<unk>}. In \secref{sec:implementation} we describe how we empirically derive the vocabulary $V$ and in \secref{sec:model} we explain how the copy mechanism helps in overcoming the unknown tokens problem.

\figref{fig:abstraction} shows the output of this process. The original class is presented in \listref{lst:fooOriginal} and \listref{lst:buggyContext} displays the buggy class after Buggy Context Abstraction. \listref{lst:fooAbstracted} illustrates the class when tokens that are out of vocabulary are replaced with the unknown token <unk>. 
Programming language tokens such as \texttt{class} and \texttt{int} are not replaced with <unk> because they are part of the vocabulary. Other in-vocabulary tokens include common variable names such as \texttt{i}.  
Our sequence-to-sequence network receives \listref{lst:buggyContext} as input.

\subsection{Sequence-to-Sequence Network}
\label{sec:seq2seq}
In this phase we train \approach to learn how to generate a fix for a given bug. Specifically, we train a Sequence-to-Sequence Network with Encoder-Decoder model (with attention and copy mechanism) to translate the \textit{abstract buggy context} of a bug to the corresponding target fixed line $f_l$. To train such a network we rely on a large dataset of bug fixes mined from different sources, explained in \secref{sec:datasets}. The bug fixes are divided into training and testing data, which are used to train and evaluate the Sequence-to-Sequence Network described in \secref{sec:model}.

\subsubsection{Model}
\label{sec:model}

\begin{figure}
    \centering
    \includegraphics[width=0.5\textwidth]{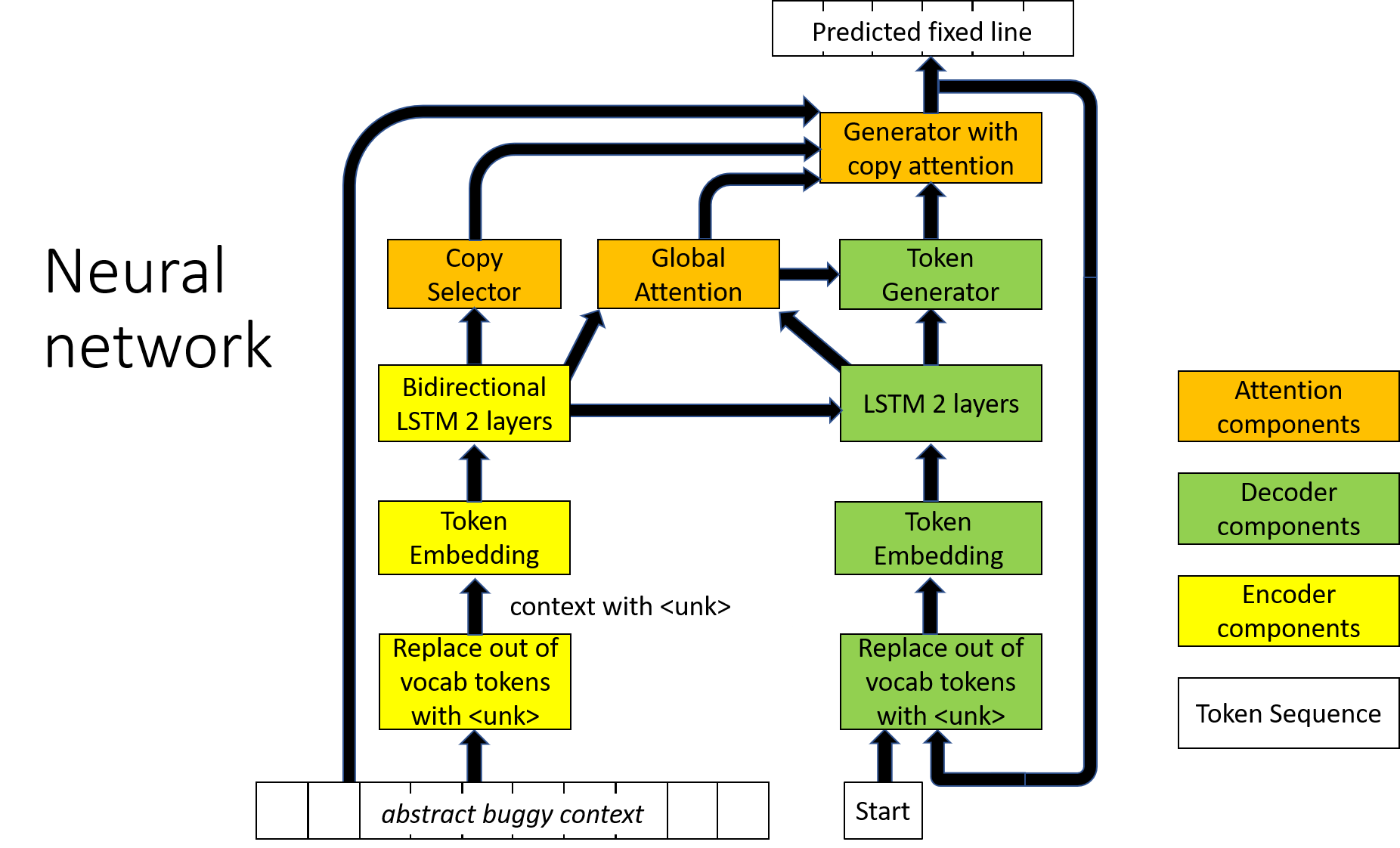}
    \caption{Sequence-to-sequence model used in \approach.}
    \label{fig:seq2seq}
\end{figure}

Figure~\ref{fig:seq2seq} shows our model for sequence-to-sequence learning to create Java source code patches. The basis of our model is a recurrent neural network similar to a natural language processing architecture \cite{sutskever2014sequence}. During training, 
the source token sequence $X = [x_1,...,x_n]$ (\ie \textit{abstract buggy context}) is provided to the encoder, where $n$ is the token length of \textit{abstract buggy context}. Then, the decoder produces the target sequence $Y = [y_1,...,y_m]$ (\ie the fixed line), where $m$ is the token length of the fixed line. Back propagation is used to update the parameters in the network with stochastic gradient decent during training \cite{kiefer1952stochastic}. The trained parameters are unchanged during inference (patch generation in our case). 

\textbf{Encoder} The encoder is a recurrent neural network using LSTM gates to process the input \cite{hochreiter1997long}. It is a bidirectional encoder which allows the encoding for a token to incorporate information from other tokens both before and after it in the input data \cite{schuster1997bidirectional}.
The encoder converts the source sequence $X = [x_1,...,x_n]$ into a sequence of encoder hidden states $h_i$ using a learnable recurrence function $g_e$. After reading the last token, the last hidden state, $h_n^e$ is used as the context vector $c$ for use in initializing the decoder \cite{cho2014learning}:
\begin{equation}
h_i^e = g_e(x_i,h_{i-1}^e); 
\end{equation}

\textbf{Decoder} The decoder is also a recurrent neural network using LSTM gates. When initialized by the encoder, it begins production of the patch candidate by receiving the special \textit{start} token as input $y_0$. For each previous output token $y_{j-1}$, the decoder updates its hidden state $h_j^d$ using the learnable recurrence function $g_d$ \cite{cho2014learning}:
\begin{equation}
h_j^d = g_d(y_{j-1},h_{j-1}^d,c)
\end{equation}

The initial value $h_0^d$ is provided by a learnable bridge function of the encoder state. The decoder states $h_j^d$ are used in for token generation by the attention and copy mechanisms in \equref{eq:attntokenprob} and \equref{eq:copyprobability}.
The decoder stops generating new tokens when the last token generated by the model is a special end-of-sequence token.

\textbf{Attention} In addition, we use an attention mechanism that provides a way to create a more specific context vector $c_j$ for each output token $y_{j}$ from the decoder using a linear combination of the hidden encoder states $h_i^e$ \cite{bahdanau2014neural}:
\begin{equation}
\label{eq:attndistribution}
c_j=\sum_{i=1}^{n}\alpha^j_{i}h_i^e
\end{equation}
Where $\alpha^j_i$ represents learnable attention weights. This context vector $c_j$ is used by a learnable function $g_a$ to allow each output token $y_{j}$ to pay "\textit{attention}" to different encoder hidden states when predicting a token from vocabulary $V$:
\begin{equation}
\label{eq:attntokenprob}
P_{V}(y_{j} \mid y_{j-1},y_{j-2},...,y_{0},c_j) = g_a(h_{j}^d, y_{j-1}, c_j)
\end{equation}

\textbf{Copy} In \secref{sec:nmt} we presented the intuition behind the copy mechanism, while in this section we describe how it operates during patch generation. The copy mechanism can significantly improve the performance of the system by allowing the model to select a token from any of the tokens provided in the \textit{abstract buggy context}, even when the tokens are not contained in the training vocabulary. We empirically show the improvements offered by this approach by comparing it to the vanilla  sequence-to-sequence model without a copy mechanism in \secref{sec:answer-rq2}.
The copy mechanism contributes to \equref{eq:attntokenprob} to produce a token candidate. This component calculates $p_{gen}$, the probability that the decoder generates a token from its initial vocabulary. And $1-p_{gen}$ is the probability to copy a token from the input tokens depending on the attention vector $\alpha^j$ in \equref{eq:attndistribution} \cite{SeeLM17}:
\begin{equation}
\label{eq:copyprobability}
p_{gen} = g_c(h_j^d,y_{t-1},c_j)
\end{equation}
\begin{equation}
\label{eq:copyextended}
P(y_j) = p_{gen}P_{V}(y_j)+(1-p_{gen})\sum_{i:x_i=y_j}a^j_i
\end{equation}
$g_c$ in \equref{eq:copyprobability} is learnable function. Using \equref{eq:copyextended}, the output token $y_j$ for the current decoder state is selected from the set of all tokens that are either:
\begin{enumerate*}
\item tokens in the training vocabulary  (including the <unk> token) or
\item tokens in the \textit{abstract buggy context}.
\end{enumerate*} Although there are no <unk> targets in the training set for patches, if the $P_V$ computation is very uncertain which token is correct, it may happen to have a high likelihood for <unk>. If at the same time, $p_{gen}$ is high then a <unk> token will be produced as the copy mechanism did not replace it. Such outputs are discarded as discussed in \secref{sec:patch-preparation}.

\subsection{Patch Inference}
\label{sec:patch-inference}

Once the sequence-to-sequence network is trained, it can be used to generate patches for projects outside of the training dataset. During patch inference, we still generate \textit{abstract buggy context} for the bug, as described in \secref{sec:context-abstraction}. But we will use beam search to generate multiple likely patches for the same buggy line, as done in related work \cite{tufano2018arxiv,ahmed2018compilation}. Beam search works by keeping the $n$ best sequences up to the current decoder state. The successors of these states are computed and ranked based on their cumulative probability; and the next $n$ best sequences are passed to next decoder state. $n$ is often called the  width or beam size, and beam search with an infinite $n$ corresponds to doing a complete breath-first-search. 
In \listref{lst:withCopy}, we have an example of predictions with beam size 5 for the bug presented in \listref{lst:buggyContext}. Each row is one prediction from the model, representing one potential bug fix, and each of them is further processed by the patch preparation step described below.

\subsection{Patch preparation}
\label{sec:patch-preparation}

\begin{figure}
\begin{center}
\begin{minipage}[t]{.33\linewidth}
\begin{lstlisting}[columns=flexible,frame=single,basicstyle=\footnotesize,label={lst:withoutCopy},caption={Without copy mechanism},captionpos=b,breaklines=true,numbers=left, numbersep=6pt]
return 1 ;
return i ;
return <unk> ;
return <unk> + 1 ;
return <unk> . <unk>;
\end{lstlisting}
\end{minipage}
\hfill
\begin{minipage}[t]{.28\linewidth}
\begin{lstlisting}[columns=flexible,frame=single,basicstyle=\footnotesize,label={lst:withCopy},caption={Network output},captionpos=b,breaklines=true]
return 1 ;
return i ;
return <unk> ;
return bar + 1 ;
return Foo . bar ;
\end{lstlisting}
\end{minipage}
\hfill
\begin{minipage}[t]{.3\linewidth}
\begin{lstlisting}[columns=flexible,frame=single,basicstyle=\footnotesize,label={lst:patchprep},caption={After patch preparation},captionpos=b,breaklines=true]
return 1;
return i;
// discarded
return bar+1;
return Foo.bar;
\end{lstlisting}
\end{minipage}
\end{center}
\caption{Patch preparation step using copy mechanism}
\label{fig:modelPrediction}
\end{figure}

The raw output from the sequence-to-sequence network cannot be used as a patch directly.
First, the predictions might still contain <unk> tokens not handled by the copy mechanism. \listref{lst:withoutCopy} illustrates token values before the copy mechanism replaces <unk> for samples 4 and 5. But the copy mechanism may not replace all such tokens as seen in sample 3 of \listref{lst:withCopy}.
Second, the predictions contain a space between every token, which is not well-formed source code in many cases. (For example, a space is not allowed between the dot separator, ".", and a method call, but a space is required between a type and the corresponding identifier name.)

Consequently, we have a final patch preparation step as follows.
We discard all line predictions that contain <unk> and we reformulate the remaining predictions into well-formed source code by removing or adding the required spaces. An example is shown between \listref{lst:withCopy} and \listref{lst:patchprep}, whitespaces are adjusted and the third prediction from \listref{lst:withCopy} is removed since it contains <unk> token.
Each one of the line predictions is used to create a candidate program by replacing the original buggy line $b_{i}^{l}$ (\ie the <START\_BUG>, <END\_BUG> and all tokens in between are replaced with the model output).

More formally,
the remaining candidate fixed lines, $cand_i = \{pre_i^1,pre_i^2,..\}$, will replace the buggy line $b_i^l$ in buggy system $b^s$ and generate candidate patches $\{patch_i^1,patch_i^2,...\}$, which should be verified with any patch validation technique, such as test suite validation. When the test suite is weak to specify the bug, we can have different patches $\{patch_i^1, patch_j^1,...\}$ for different bug locations $\{l_i,l_j,...\}$ that passed the test suite. Then, the correctness can be verified, for example, by manual inspection.

\subsection{Implementation Details \& Parameter Settings}
\label{sec:implementation}

\textbf{Library.} We have implemented our Encoder-Decoder model using OpenNMT-py \cite{opennmt}, built in the Python programming language and the PyTorch neural network platform \cite{paszke2017automatic}.

\textbf{Vocabulary} In this paper, we consider a vocabulary of the 1,000 most common tokens.
To the best of our knowledge, this is one of the largest vocabularies considered for machine learning for patch generation:
for comparison, DeepFix \cite{gupta2017deepfix} has a vocabulary size of 129 words, and Tufano \etal \cite{tufano2018arxiv} considered a vocabulary size of 430 words.

\textbf{Limit for truncation} We truncate if the \textit{abstract buggy context} is longer than 1,000 tokens. This is motivated by \figref{fig:FileTokens}, where we can see that \textit{abstract buggy context} is often less than 1,000 tokens long. \approach truncates by keeping the buggy line but removing statements, class definitions, and method definitions until \textit{abstract buggy context} is 1,000 tokens or less.

\textbf{Network parameters} 
We explored a variety of settings and network topologies for \approach. Most major design decisions are verified with ablation experiments that change a single variable at a time as detailed further in \secref{sec:ablation}. We train our model with a batch size of 32 for 10,000 iterations. To prevent overfitting, we use a dropout of 0.3. In relation to the components shown in \figref{fig:seq2seq}, below are the primary matrix sizes associated with each component along with a reference to the equations in \secref{sec:model} to which they relate:
\begin{itemize}
    \item Token embedding (our model uses the same embedding for both $g_e$ and $g_d$): 1,004x256 (1,000 + 4 special tokens)
    \item Encoder bidirectional LSTM (part of $g_e$ fuction): 256x256x4x2x2
    \item Decoder LSTM (part of $g_d$ function): 512x256x4x2 + 256x256x4x2
    \item Token generator (part of $g_a$ function): 256x1004
    \item Bridge between encoder and decoder (path for $h_i^e$ to initialize $h_0^d$): 256x256x2
    \item Global Attention ($\alpha_i^j$ weights): 256x256 + 512x256
    \item Copy selector ($g_c$ function): 256x1
\end{itemize}

We use a beam size of 50 during inference, which is the default value used in the literature \cite{tufano2018arxiv}\cite{ahmed2018compilation} and which proves to be good empirically.

\textbf{Input and output summary}
The input to SequenceR is a Java class of any size. The non-empty faulty line within a method on which to attempt repair has been identified by another technique (usually line-based fault localization). The output is the fixed line which must have fewer than 100 tokens with our current model.

\textbf{Usage}
After \approach is trained, we can use it to predict fixes to a bug. \approach takes as input the buggy file and a line number indicating where the bug is. The output is a list of patches in the diff format, so that the user can run their own patch validation step, which could either be test validation or manual inspection.

The source code of \approach is available at \url{https://github.com/kth/SequenceR}, together with the best model we have identified and the synthesized patches.

\section{Evaluation}
\label{sec:evaluation}

In this section, we describe our evaluation of \approach.

\subsection{Research Questions}

The two first research questions focus on machine learning:
\begin{itemize}
    \item RQ1: To what extent can the fixed line be perfectly predicted?
    \item RQ2: 
     How often does the copy mechanism generate out-of-vocabulary tokens for a patch, and which parts of \textit{abstract buggy context} are referenced for the copy?
\end{itemize}

The last two research questions look at the system from a domain-specific perspective: we assess the performance of \approach from the viewpoint of program repair research.
\begin{itemize}
    \item RQ3: How effective is \approach's sequence-to-sequence learning in fixing bugs in the well-established Defects4J benchmark?
    \item RQ4: What repair operators are captured with sequence-to-sequence learning?
\end{itemize}

\subsection{Experimental Methodology}

\subsubsection{Methodology for RQ1}

We train \approach with the parameter settings described in \secref{sec:implementation}. The training and validation accuracy and perplexity will be plotted. Perplexity (ppl) is a measurement of how well a model predicts a sample and is defined as:
$$ppl(X,Y) = exp(\frac{-\sum_{i=1}^{\mid Y \mid} \log P(y_{i} \mid y_{i-1},\ldots,y_{1},X)}{\mid Y \mid})$$
where $X$ is the source sequence, $Y$ is the true target sequence and $y_{i}$ is the \textit{i}-th target token \cite{opennmt}. Luong \etal found a strong correlation between a low perplexity value and high translation quality \cite{luong2014addressing}.

The resulting model is tested on our testing dataset, CodRep4 (see \secref{sec:dataprep}). Next, in order to compare \approach against the state-of-the-art approach by Tufano \etal \cite{tufano2018arxiv}, we created CodRep4Medium. It is a subset of CodRep4 containing 1,116 samples where the buggy method length is limited to 100 tokens.

\subsubsection{Methodology for RQ2}
\label{subsubsec:RQ2_method}
To evaluate the effectiveness of the copy mechanism (described in \secref{sec:model}), we consider all samples from CodRep4. For each successfully predicted line, we categorize tokens in that line based on whether the token is in the vocabulary or not. And at the same time, for tokens that are out-of-vocabulary but are copied from the input sequence, we try to find the original location of the copied token.
By analyzing the original location of out-of-vocabulary tokens,  we can measure the importance of the context, in particular of the \textit{abstract buggy context} we define in this paper. The copy mechanism allows the system to be more powerful by providing more tokens beyond the vocabulary to be used in the patch.

\subsubsection{Methodology for RQ3}
\label{sec:RQ3_method_defects4J}

We evaluate \approach on Defects4J \cite{just2014defects4j}, which is a collection of reproducible Java bugs. Most recent approaches in program repair research on Java use Defects4J as an evaluation benchmark \cite{martinez2017automatic,xiong2017precise,xin2017leveraging,wen2018context,jiang2018shaping}.

Since the scope of our paper is on one-line patches, we first focus on Defects4J bugs that have been fixed by developers by replacing one single line (there are 75 such bugs).
In order to study the effectiveness of sequence-to-sequence itself, we isolate the fault localization step as follows: the input to \approach is the actual buggy file and the buggy line number. \approach then produces a list of patches (recall that beam search produces several candidate patches). 
All patches are compiled and then executed against the test suite written by the developer. 

Each candidate patch generated by \approach is then categorized as follows:
\begin{itemize}
\item \textbf{Compilable patch}: The patch can be compiled.
\item \textbf{Plausible patch}: The patch is compilable and passes the test suite. The patch may yet be incorrect because of the overfitting problem \cite{smith2015cure}.
\item \textbf{Correct patch}: The patch passes the test suite, and is semantically equivalent to the human patch. We hand-check for semantic equivalence for this evaluation.
\end{itemize}

As per the definitions, there is a strict inclusion structure in those categories: correct patches are necessarily plausible and compilable, plausible patches are necessarily compilable. 

\subsubsection{Methodology for RQ4}
For RQ4, we aim at having a qualitative understanding of the cases for which our sequence-to-sequence repair approach works. 
This research question is motivated by the need to understand what grammatically correct code transformations are captured by \approach, even though it is purely a token-based approach with no first class AST or grammar knowledge. 
For gaining this understanding, we use a mixed method combining grounded theory and targeted analysis. The results would be an understanding of the variety of repair operators and programming language syntax captured by \approach in cases where the model output correctly matches the test data.
For the grounded theory, we have been regularly sampling successful cases, \ie cases in our testing dataset CodRep4 for which \approach was able to predict the fixed line, for each case, the authors reached a consensus to know whether 1) the case is interesting from a programming perspective (\eg it represents a common bug fix pattern), and 2) the case highlights a phenomenon that has already been covered in a previously found case.
For the targeted analysis, we specifically searched for 2 kinds of results: cases where the copy mechanism was used and cases where a specific programming construct was involved (method call, field reference and string literals).

\subsection{Training Data}
\label{sec:datasets}

\approach is trained based on past modifications made to source code, \ie it is trained on past commits.
In our experiments, we combine two sources of past commits,
the CodRep dataset \cite{CodRep18} and the Bugs2Fix dataset \cite{tufano2018arxiv}, into what appears to be the largest dataset of one-line bug fixes published to date. Both datasets 1) consider Java code and 2) have been built based on the history of open-source projects.

The CodRep dataset focuses solely on one-line source code fixes (aka one-line patches), it contains 5 datasets curated from real commits on open-source projects. The Bugs2Fix dataset contains diffs mined from Github between March 2010 and October 2017 for bug-fixing commits (based on heuristics to only consider bug-fixing commits). Neither dataset requires the buggy project to have a test suite for exposing the buggy behavior, instead they are focusing on collecting bug fix commits.

\subsubsection{Data Preparation}
\label{sec:dataprep}

Since CodRep and Bugs2Fix datasets are in different formats, we first unify these two datasets as follows. First, we only keep diffs from Bugs2Fix which are fixes with a single line replacement. Further, we filter out certain diffs if the changes are outside of a method.

Since the Bugs2Fix dataset comes from a generic bug-fix data mining which includes multi-line fixes and fixes outside of methods, we can look at its statistics to help understand the generality of \approach. Bugs2Fix contains 92,849 commits. 15,548 of these (17\%) are one-line patches within a method, and are within the problem domain of \approach. 

After preparing the dataset, we divide it into training and testing data. CodRep is originally split into 5 parts, numbered from 1 to 5, with each part containing commits from different groups of projects. Our training data consists of CodRep datasets 1,2,3 \& 5 and the Bugs2Fix dataset. Our testing data is CodRep dataset 4 (or CodRep4 for short). We chose dataset 4 because it is approximately 20\% of the entire CodRep data (data set 1 is less than 10\% and data set 5 is over 30\%) and because CodRep 4 contains a broad and representative set of projects on which to evaluate \cite{CodRep18}.

Furthermore, we ensure there are no duplicate samples between the training and testing datasets.
During the model setup, we use a random subset of 95\% of the training data for model training and 5\% as our validation dataset.

\subsubsection{Descriptive Statistics of the Datasets}

In total, we have 35,578 samples in our training set and 4,711 samples in our testing set.

\begin{figure}
\begin{center}
\includegraphics[width=.45\textwidth]{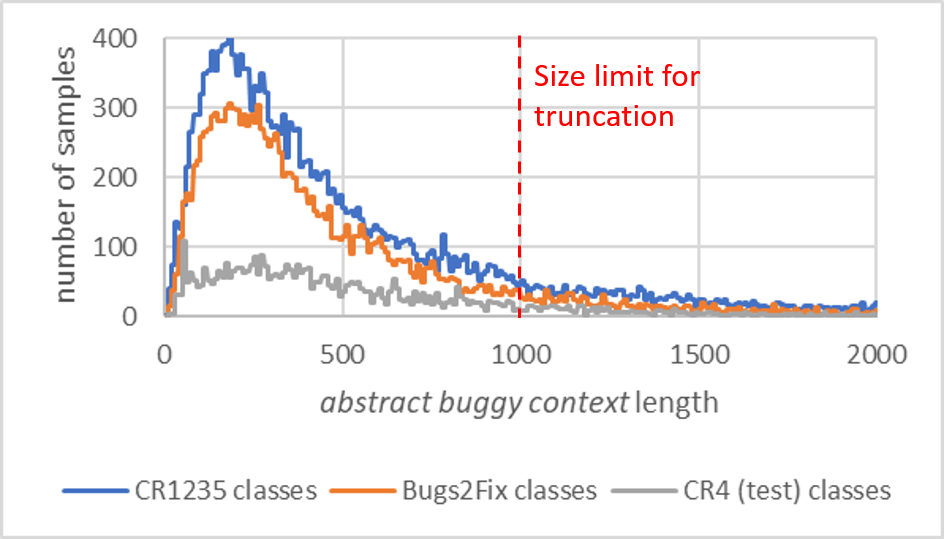}
\caption{Only 14\% of samples exceed the 1K token length limit and require truncation.}
\label{fig:FileTokens}
\end{center}
\end{figure}

\textbf{Input Size}
Figure~\ref{fig:FileTokens} shows the size distribution of the \textit{abstract buggy context} in number of tokens before truncation is done. The CodRep training data has a median token length of 372; the Bugs2Fix dataset has a median length of 340 tokens; and the testing dataset has a median length of 411. These variations are a result of using different Java projects in the datasets, but we observe that the distribution of lengths is similar.

\begin{figure}
\begin{center}
\includegraphics[width=.45\textwidth]{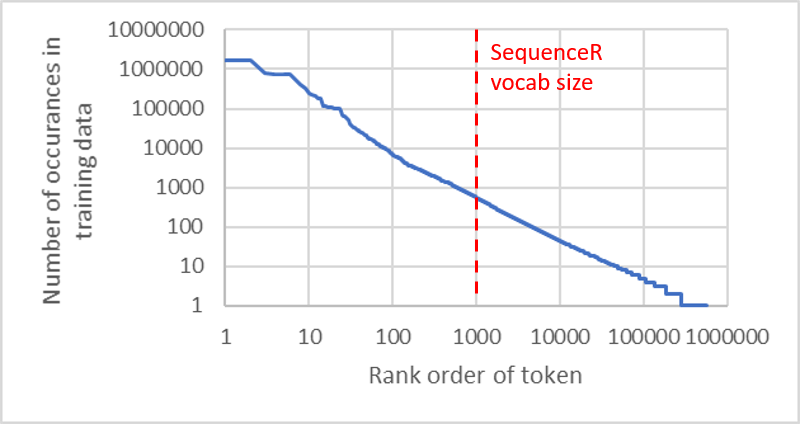}
\caption{Overview of vocabulary: token count occurrences follow a Zipf's law distribution.}
\label{fig:TokenDist}
\end{center}
\end{figure}

\textbf{Prediction Size}
The lines from the \textit{abstract buggy context} samples in our dataset had a median length of 6. 99\% of the lines were 30 tokens or fewer, which fits well typical output sizes used for natural language processing.
To sum up, the order of magnitude of the sequence-to-sequence prediction receives an input sequence with an average length of 350 tokens and produces an output sequence with an average length of 6 tokens.

\textbf{Vocabulary Size}
In our training data, the full vocabulary is 567,304 different tokens. \autoref{fig:TokenDist} shows the distribution of the number of occurrences for the whole vocabulary. It is a typical power-law like distribution with a long tail. We limit our training vocabulary to the 1,000 most common tokens.

\subsection{Experimental Results}

\subsubsection{Answer to RQ1: Perfect Predictions}

\begin{table}
\resizebox{\columnwidth}{!}{%
\begin{tabular}{|l|l|l|}
\hline
Approach & \multicolumn{2}{|c|}{Prediction Accuracy} \\ \cline{2-3}
& CodRep4Medium & CodRep4  \\ \hline
simple seq2seq line2line, no copy & 77/1116 (6.9\%) & 206/4711 (4.4\%) \\ \hline
Tufano \etal \cite{tufano2018arxiv} & 157/1116 (14.1\%) &  N/A \\ \hline
\approach & 344/1116 (30.8\%) & 950/4711 (20.2\%)  \\ \hline
\end{tabular}
}
\caption{Comparison with state-of-the-art approach by Tufano \etal}
\label{tab:rq1table}
\end{table}

We trained our model on a GPU (Nvidia K80) for 1.2 hours. For a typical training run on our golden model, \figref{fig:accuracy} shows the training and validation accuracy per token generated (the accuracy for the entire patch would be lower) and \figref{fig:ppl} shows the perplexity (ppl) per token generated over the training and validation datasets. In this particular run, the best results for both the perplexity and accuracy on the validation dataset occur at 10,500 iterations. We chose 10,000 iterations as the standard training time for our model.

\textbf{CodRep4} On the 4,711 prediction tasks of our best model, \approach is able to generate the perfect fix in 950 cases (from \tabref{tab:rq1table}). In all those cases, the predicted line that replaces the buggy line is exactly the line fix implemented by the developer. The copy mechanism is used in a number of cases, this will be further discussed in \autoref{sec:answer-rq2}. 

\textbf{Comparison to state-of-the-art}
To the best of our knowledge, the state-of-the-art approaches are from Tufano \etal \cite{tufano2018arxiv}  and Hata \etal \cite{Hata2018}. We only compare against Tufano \etal since their approach has been open sourced while that one of Hata \etal was not made available at the time of writing this paper. The approach used by Tufano \etal is limited to fixes only inside small methods, consisting of less than 100 tokens. The limitation is due to the fact that their approach generates the entire fixed source code method as output of the decoder. This means that the decoder may need to generate a long sequence of source code tokens, which is one of the major challenges for NMT models \cite{DBLP:journals/corr/KoehnK17}.
\approach does not make any assumption on the size of the buggy method. In order to compare against \cite{tufano2018arxiv}, we select those 1,116 tasks from CodRep4 where the buggy line resides in a method smaller than 100 tokens. Those 1,116 tasks are called the CodRep4Medium testing dataset.

Our testing accuracy for both CodRep4 and CodRep4Medium are shown in \tabref{tab:rq1table}. From the table, we see that the accuracy of \approach is 344/1,116 (30.8\%) while Tufano \etal \cite{tufano2018arxiv} is 157/1,116 (14.1\%).
This is a clear indicator that \approach outperforms the current state-of-the-art showing twice as many correct predictions. 
It shows that our construction of the \textit{abstract buggy context}, together with the copy mechanism, leads to higher accuracy than only having the buggy method as context with a specific encoding for variables. Recent fault localization research \cite{ZouLXEZ2019} indicates that best-in-class techniques can predict the faulty line 44\% of the time and the faulty method 68\% of the time. If we extrapolate these percentages to our data, \approach is more likely to find correct one-line patches than the prior work \cite{tufano2018arxiv} is to find method replacements, and \approach can process and repair larger methods as demonstrated by the right-hand column of \tabref{tab:rq1table}.

We now concentrate on the effectiveness of the approach depending on the buggy method length.
Overall, we observe that \approach has a lower accuracy on longer methods (30.8\% accuracy on CodRep4Medium, 20.2\% accuracy on CodRep4). This phenomenon is explained by the fact that fixes in long methods are usually more complex and involve more context variables, identifiers and literals that are not easily captured by the learning system. This phenomenon has also been previously observed \cite{tufano2018arxiv}.

\begin{samepage}
\begin{figure}
\begin{center}
\includegraphics[width=.4\textwidth]{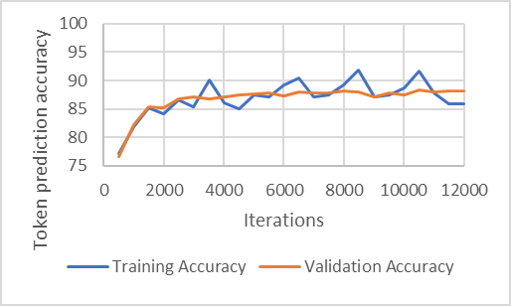}
\caption{Training and validation accuracy}
\label{fig:accuracy}
\end{center}
\begin{center}
\includegraphics[width=.4\textwidth]{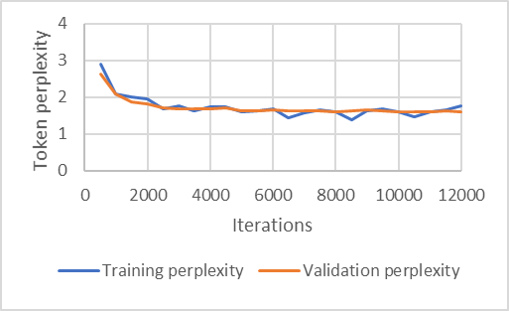}
\caption{Training and validation perplexity}
\label{fig:ppl}
\end{center}
\end{figure}
\end{samepage}

\subsubsection{Answer to RQ2: Copy Mechanism}
\label{sec:answer-rq2}

\begin{figure}
    \centering
    \includegraphics[width=\columnwidth]{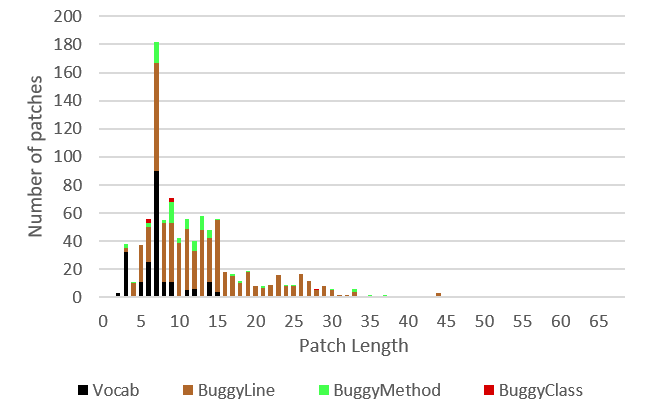}
    \caption{Histogram showing correctly generated patches: 1) that only use tokens in our 1,000 token vocabulary, 2) that need to copy tokens from the buggy line, 3) from the buggy method and 4) from the buggy class.}
    \label{fig:VocabHist}
\end{figure}

We now look at to what extent the copy mechanism is used. 
\figref{fig:VocabHist} shows the origin of tokens in successfully predicted lines, per patch size.
Let us consider the highest bar, corresponding to all successfully predicted lines consisting of 7 tokens. For those 7-token patches, the black bar means that all tokens are taken from the vocabulary. The non-black bars mean that the copy mechanism has been used to predict the line fix. Overall, there is a minority of patches (216/950, 23\%) for which all tokens come from the vocabulary. At the extreme, the longest successful patch generated by \approach was 68 tokens long, but the longest successful patch without the copy mechanism was only 27 tokens long.

\figref{fig:VocabHist} also lets us analyze the location origin of the copied token. The brown bars represent those patches for which copied tokens all come from the buggy line: this is the majority of cases (641/950, 68\%). However, we also observe cases where some copied tokens have been taken from the buggy method (green bars) and cases where the copied tokens has been taken from the buggy class (red bars), \ie taken from the class context as captured in our encoding. 

As an example, \listref{lst:dataflow} replaces variable \texttt{masterNode} with \texttt{nonMasterNode} as in the correct human patch. \texttt{nonMasterNode} in the fixed line does not occur in our training data and hence it is not in our 1000 token vocabulary. Therefore, \approach was able to generate this patch because it copied the out-of-vocabulary token \texttt{nonMasterNode} from within the buggy method.
As this example is a 4 token long patch, it would contribute to the green bar for patch length 4 in \autoref{fig:VocabHist}.

\noindent
\begin{minipage}{\linewidth}
\begin{lstlisting}[language=diff,columns=flexible, frame=single, basicstyle=\footnotesize,label={lst:dataflow},caption={Example of the copy mechanism creating a correct patch by incorporating a variable which is not in the vocabulary from the broader context around the buggy line.},captionpos=b,breaklines=true]
while( nonMasterNode == null ) {
  nonMasterNode=randomFrom( internalCluster().getNodeNames());
  if( nonMasterNode.equals( masterNode ) ) {
-      masterNode = null;
+      nonMasterNode = null;
  }
}
\end{lstlisting}
\end{minipage}

Overall, \figref{fig:VocabHist} shows that the copy mechanism is extensively used (734/950, 77\%) and that our class level abstraction enables us to predict difficult cases where only the buggy line or the buggy method would not have been enough.

In order to understand the benefits of context size with the copy mechanism, we measured the distance in tokens to reach a copied token used to generate a patch. In the 87 cases where a copied token was needed from the buggy method $b_m$, the median distance from the buggy line $b_l$ to the nearest use of the copied token was 9 tokens, 90\% of the 87 cases were within 49 tokens of $b_l$, and 100\% were found within a 122 token distance. In the 7 cases when a copied token was needed from the buggy class $b_c$, the median distance to the copied token from $b_m$ was 25 tokens, and 100\% were found within a 241 token distance. In addition to ablation study results discussed is \secref{sec:ablation}, the preceding data supports our decision to create the \textit{abstract buggy context}. 

\subsection{Answer to RQ3: Defects4J Evaluation}
\label{sec:answer-to-rq3}

\begin{figure}
\begin{center}
\begin{tikzpicture}[x={(0.1,0)},scale=0.6]
\foreach  \l/\x/\c[count=\y] in {
with correct patches/14/green,
with plausible patches/19/blue,
with compilable patches/53/yellow,
Bugs with patches/58/orange,
Total bugs/75/gray}
{\node[left] at (0,\y/2) {\scriptsize\l};
\fill[\c] (0,\y/2-.2) rectangle (\x,\y/2+.2);
\node[right] at (\x, \y/2) {\x};}
\draw (0,0) -- (90,0);
\foreach \x in {0, 15, ..., 90}
{\draw (\x,.2) -- (\x,0) node[below] {\x};}
\draw (0,0) -- (0,5.5/2);
\end{tikzpicture}
\end{center}
\caption{\approach results on the 75 one-line Defects4J bugs.}
\label{fig:bugStat}
\end{figure}
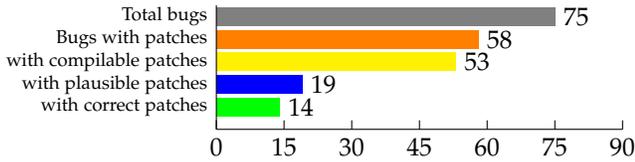

As explained in \secref{sec:RQ3_method_defects4J}, we consider 75 Defects4J bugs that have been fixed with a one-line patch by human developers. In total \approach finds 2,321 patches for 58 of the 75 bugs. The main reason that we are unable to fix the remaining 17 bugs is due to fact that some bugs are not localized inside a method, which is a requirement for the fault localization step that \approach assumes as input. 
\listref{lst:notInMethod} is one such example where the Defects4j bug is not localized inside a method.
We have 2,321 patches instead of 2,900 (58x50) because some predictions are filtered by the patch preparation step (\secref{sec:patch-preparation}), \ie patches that contain the <unk> token.
The statistics about all bugs can be found in \figref{fig:bugStat}.
Out of 75 bugs, \approach successfully generated at least one patch for 58 bugs, 53 bugs have at least one compilable patch, 19 bugs have at least one patch that passed all the tests (\ie are plausible) and 14 bugs are considered to be correctly fixed (semantically identical to the human-written patch). Of these 14 bugs, in 12 cases the plausible patch with the highest ranking in the beam search results was the semantically correct patch.

\begin{lstlisting}[language=diff,columns=flexible, frame=single, basicstyle=\footnotesize,label={lst:notInMethod},caption={An example of Defects4J defect (Math 104) where the bug is not localized inside a method. In this case, a class variable is changed.},captionpos=b,breaklines=true]
- private static final double DEFAULT_EPSILON = 10e-9;
+ private static final double DEFAULT_EPSILON = 10e-15;
\end{lstlisting}

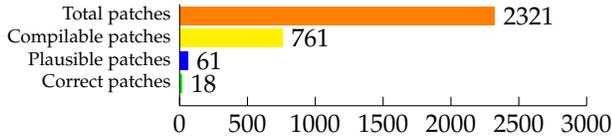
\begin{figure}
\begin{center}
\begin{tikzpicture}[x={(.003,0)},scale=0.6]
\foreach  \l/\x/\c[count=\y] in {
Correct patches/18/green,
Plausible patches/61/blue,
Compilable patches/761/yellow,
Total patches/2321/orange}
{\node[left] at (0,\y/2) {\scriptsize\l};
\fill[\c] (0,\y/2-.2) rectangle (\x,\y/2+.2);
\node[right] at (\x, \y/2) {\x};}
\draw (0,0) -- (3000,0);
\foreach \x in {0, 500, ..., 3000}
{\draw (\x,.2) -- (\x,0) node[below] {\x};}
\draw (0,0) -- (0,4.5/2);
\end{tikzpicture}
\end{center}
\caption{Stastistics on patches synthesized by \approach for the 75 one-line Defects4J bugs.}
\label{fig:patchStat}
\end{figure}

\figref{fig:patchStat} gives a different perspective on this data, focusing on patches (and not bugs). 
\approach is able to generate 761 compilable patches (33\% of all patches).
\approach finds 61 plausible patches spread over 19 bugs, thus there can be several plausible patches for the same bug, a phenomenon well-known in the program repair field \cite{martinez2017automatic}.
One reason is that some Defects4J bugs have a weak test suite. 
To the best of our knowledge, we are the first to report the correctness of patches generated by a sequence-to-sequence model, where correctness means passing the test suite and being semantically equivalent to the human patch.
In the end,  \approach is able to generate 18 patches that are semantically equivalent to the correct bug fix.

For \approach applied to Defects4J bugs, we observe that out of 61 plausible patches, 18 are correct, which is a ratio of 30\%. An analysis of prior techniques which used a different benchmarck in C (GenProg \cite{GenProg12}, RSRepair \cite{RSRepair14}, and AE \cite{AE13}) shows that they have a correct patch ratio of less than 12\% \cite{Qi2015}. We did not evaluate \approach on the same benchmark as this prior work (we target Java not C), but the ratio is evidence that \approach has learned to produce outputs which represent reasonable patch proposals.

Although we did not directly include fault localization in our evaluation of \approach, we can estimate the performance of a repair system which includes state-of-the-art fault localization techniques \cite{ZouLXEZ2019} as follows. It has been shown that there is an estimated 44\% success of correctly identifying a faulty line in the top 10 candidates. Hence, in order to process 75 total bugs from Defects4J, 750 candidate abstract buggy contexts would need to be prepared for input to our model. We have run fault localization with Gzoltar \cite{GZoltar12} and found that it successfully localized the faulty line for 9 of the 14 bugs for which \approach found a correct fix.  

Let us now discuss timing. We estimate the machine time required to automatically find patches for 75 bugs with the summation below\footnote{Our Defects4J testing was run on an Intel Core i7 at 3.5GHz and our sequence-to-sequence model was run on an Nvidia K80.}: 
\begin{itemize}
    \item Estimated time to run fault localization on 75 bugs and identify 10 likely faulty line locations: 112 minutes
    \item Time to create 750 \textit{abstract buggy contexts} (10 created for each bug): 29 minutes
    \item Time to create 37,500 patch candidates (50 candidates created from beam size 50 for each \textit{abstract buggy context}): 9 minutes
    \item Estimated time to prune raw patches down to 23,210 total patches: 2 minutes
    \item Time to attempt compile on 23,210 patches: 1378 minutes
    \item Time to run test cases on 7,610 patches: 6287 minutes
    \item Final result estimated to take 130 total machine hours to find patches which correctly fix 9 bugs.
\end{itemize}

\listref{lst:math75} shows the \approach patch for Math 75, which is semantically equivalent to the human patch. We observe that it contains some unnecessary parentheses, and the same behavior occasionally occurs in other patches found by \approach. We have observed unnecessary parenthesis in some of the human-generated patches in our training data and \approach occasionally replicates this human style. In this case, the parentheses do not change the order of evaluation. Therefore the \approach patch for Math 75 is semantically equivalent to the human patch.

Interestingly, \texttt{getPct} is not part of the vocabulary, and it did not appear in the buggy method. The \texttt{getPct} method is defined in the same buggy class, as captured by our \textit{abstract buggy context}.
In Defects4J, the copy mechanism is also useful to capture the right tokens to add in the patch.

\begin{lstlisting}[language=diff,columns=flexible, frame=single, basicstyle=\footnotesize,label={lst:math75},caption={Found patch for Math 75},captionpos=b,breaklines=true,escapechar=@]
- return getCumPct((Comparable<?>) v);
+ return getPct((Comparable<?>) v); // Human patch
+ return getPct(((Comparable<?> )(v))); // @\aftergroup\diffinclcolor@@\approach@ patch
\end{lstlisting}

We now compare those results against the patches found by recent program repair tools that are publicly available.
Elixir \cite{saha2017elixir}, CapGen \cite{wen2018context} and SimFix  \cite{jiang2018shaping} have reported 26, 22, 34 correctly repaired bugs for all Defects4J bugs, where the patch is identical to the human patch or claimed as correct. Of those correctly repaired bugs, 22, 19 and 17 respectively are for the 75 one-line bugs that we consider for \approach. We notice that the majority of claimed correct patches are for one-line bugs.
We observe that \approach does not fix more one-line Defects4J bugs. 

While Elixir, CapGen, and SimFix are driven with intelligent design and require substantial configuration and handcrafted rules, our goal with \approach is to be agnostic and to \emph{not} design any repair operator upfront.
For example, CapGen implements context-aware operator selection and context-aware ingredient prioritization \cite{wen2018context}. The CapGen implementation  heavily relies on code transformation tools and carefully selected algorithms/parameters/metrics. In constrast, our \approach can be considered less heavyweight. We note that the required parameter tuning in \approach can easily be performed using grid search or other meta-optimization techniques \cite{bengio2012practical}. 
To that extent, it is remarkable that such a generic approach is able to learn bug-fixing patterns and synthesizes 18 patches that are semantically equivalent to the human repair, without any static or dynamic analysis. By providing a generic approach, \approach will improve in the future as machine learning sequence-to-sequence techniques improve, and as more bug fix training data is provided. Also, since \approach learns repair operators from examples, it could be trained on less common languages (such as COBOL).

We assume perfect fault localization while other related tools ran fault localization to localize the buggy source code. Yet, different papers use different fault localization algorithms, implementations, and granularity (\eg methods versus line). Liu \etal pointed out that because of different assumptions about fault localization, it is hard to compare different repair techniques \cite{liu2018you}. By assuming perfect fault localization, we purely focus on the patch generation step of the algorithm.

\subsection{Answer to RQ4: Qualitative Case Studies}
\label{sec:rq4answer}

We now present the diversity of repair operators that are captured by \approach. These cases are culled from the 950 correct patches \approach generated for the CodRep4Full test dataset. Both the buggy line that was part of the input is shown and the correct patch which includes examples of repair operators.
We also highlight again the effectiveness of the copy mechanism by using a \textbf{\underline{bold underlined}} font for those tokens that were copied (\ie that are outside the vocabulary of the 1,000 most common tokens). 

\subsubsection{Case study: method call change}
Our training and evaluation data consist of object-oriented Java software.
We observe that \approach captures different kinds of operations related to method calls.

\textbf{Call change}
Here a call to method \underline{\textbf{writeUTF}} is replaced by a call to method \underline{\textbf{writeString}}.
\begin{lstlisting}[language=diff,columns=flexible, frame=single, basicstyle=\footnotesize,label={lst:methodcallchange},caption={Call change},captionpos=b,breaklines=true,escapechar=@]
- out.@\aftergroup\diffremcolor@@\bfseries\underline{writeUTF}@( @\aftergroup\diffremcolor@@\bfseries\underline{failure}@ ); 
+ out.@\aftergroup\diffinclcolor@@\bfseries\underline{writeString}@( @\aftergroup\diffinclcolor@@\bfseries\underline{failure}@ );
\end{lstlisting}

\textbf{Call deletion}
The buggy line chains two method calls; this successful prediction consists of deleting one of them.

\begin{lstlisting}[language=diff,columns=flexible, frame=single, basicstyle=\footnotesize,label={lst:methoddelete},caption={Call deletion.},captionpos=b,breaklines=true,escapechar=@]
- @\aftergroup\diffremcolor@@\bfseries\underline{FieldMappers}@ x = context.@\aftergroup\diffremcolor@@\bfseries\underline{mapperService}@().@\aftergroup\diffremcolor@@\bfseries\underline{smartNameFieldMappers}@( fieldName );
+ @\aftergroup\diffinclcolor@@\bfseries\underline{FieldMappers}@ x = context.@\aftergroup\diffinclcolor@@\bfseries\underline{smartNameFieldMappers}@( fieldName );
\end{lstlisting}

\textbf{Argument addition}
In this patch, \approach adds an argument (which in Java, means calling another method).
\begin{lstlisting}[language=diff,columns=flexible, frame=single, basicstyle=\footnotesize,label={lst:methodparamchange},caption={Argument addition},captionpos=b,breaklines=true,escapechar=@]
- @\aftergroup\diffremcolor@@\bfseries\underline{stage}@.@\aftergroup\diffremcolor@@\bfseries\underline{getViewport}@().update( width, height );
+ @\aftergroup\diffinclcolor@@\bfseries\underline{stage}@.@\aftergroup\diffinclcolor@@\bfseries\underline{getViewport}@().update( width, height, true );
\end{lstlisting}

\textbf{Target change}
In this successful case, the patch also calls method \underline{\textbf{isTerminated}} but on another target (\underline{\textbf{scheduledExecutorService}} instead of \underline{\textbf{executorService}}, which is copied from the input context).
\begin{lstlisting}[language=diff,columns=flexible, frame=single, basicstyle=\footnotesize,label={lst:targetchange},caption={Target change},captionpos=b,breaklines=true,escapechar=@]
- if( !( @\aftergroup\diffremcolor@@\bfseries\underline{executorService}@.@\aftergroup\diffremcolor@@\bfseries\underline{isTerminated}@() ) ){
+ if( !( @\aftergroup\diffinclcolor@@\bfseries\underline{scheduledExecutorService}@.@\aftergroup\diffinclcolor@@\bfseries\underline{isTerminated}@() ) ){
\end{lstlisting}

\subsubsection{Case study: if-condition change}
\approach can change if conditions, and in this particular case, removes two clauses from the boolean formula. 
\begin{lstlisting}[language=diff,columns=flexible, frame=single, basicstyle=\footnotesize,label={lst:ifcondition},caption={if-condition change},captionpos=b,breaklines=true,escapechar=@]
- if( ( ( t >= 0 ) && ( t <= 1 ) ) && ( @\aftergroup\diffremcolor@@\bfseries\underline{intersection}@ != null ) )
+ if( @\aftergroup\diffinclcolor@@\bfseries\underline{intersection}@ != null )
\end{lstlisting}

\subsubsection{Case study: Java keyword change}
\approach is also able to generate patches involving the replacement of programming language keywords, indicating clues of syntax understanding.
\begin{lstlisting}[language=diff,columns=flexible, frame=single, basicstyle=\footnotesize,label={lst:javakeyword},caption={Java keyword 
change},captionpos=b,breaklines=true]
- break ;
+ continue ;
\end{lstlisting}

\subsubsection{Case study: change from field access to method call}
A good practice of software engineering is to implement encapsulation by calling methods instead of directly accessing fields, this is handled by \approach as follows (\texttt{size} to \texttt{size()})
\begin{lstlisting}[language=diff,columns=flexible, frame=single, basicstyle=\footnotesize,label={lst:field2method},caption={change from field access to method call},captionpos=b,breaklines=true,escapechar=@]
- @\aftergroup\diffremcolor@@\bfseries\underline{app}@.log( @\aftergroup\diffremcolor@@\bfseries\underline{"PixmaPackerTest"}@, ( @\aftergroup\diffremcolor@@\bfseries\underline{"Number of textures: "}@ + ( @\aftergroup\diffremcolor@@\bfseries\underline{atlas}@.@\aftergroup\diffremcolor@@\bfseries\underline{getTextures}@().size ) ) );
+ @\aftergroup\diffinclcolor@@\bfseries\underline{app}@.log( @\aftergroup\diffinclcolor@@\bfseries\underline{"PixmaPackerTest"}@, ( @\aftergroup\diffinclcolor@@\bfseries\underline{"Number of textures: "}@ + ( @\aftergroup\diffinclcolor@@\bfseries\underline{atlas}@.@\aftergroup\diffinclcolor@@\bfseries\underline{getTextures}@().size() ) ) );
\end{lstlisting}

\subsubsection{Case study: off-by-one repair}

Finally, \approach is also able to repair classical off-by-one errors.
\begin{lstlisting}[language=diff,columns=flexible, frame=single, basicstyle=\footnotesize,label={lst:offbyone},caption={off-by-one repair},captionpos=b,breaklines=true,escapechar=@]
- @\aftergroup\diffremcolor@@\bfseries\underline{nextIndex}@ = @\aftergroup\diffremcolor@@\bfseries\underline{currentIndex}@;
+ @\aftergroup\diffinclcolor@@\bfseries\underline{nextIndex}@ = ( @\aftergroup\diffinclcolor@@\bfseries\underline{currentIndex}@ ) - 1;
\end{lstlisting}

Overall, \approach uses all three kinds of token operations:
\begin{enumerate*}
    \item Token deletion, \eg \listref{lst:methoddelete};
    \item Token addition, \eg \listref{lst:methodparamchange};
    \item Token replacement, \eg \listref{lst:methodcallchange}.
\end{enumerate*}

\section{Ablation Study}
\label{sec:ablation}

We perform an ablation study to understand the relative importance of each component of our approach.
The process is as follows. First, we identify the golden model based on a greedy optimization in the parameter search space. This is the model that we described in \autoref{sec:evaluation}. Then we change one single parameter to a different reasonable value and report the performance on the same testing dataset. The ablation results demonstrate that parameter selections for the golden model produce the highest acceptance rates for the configurations we tested. The model parameters we found with our dataset are likely to yield reasonable results when training for other computer languages so long as a form of \textit{abstract buggy context} can be done to provide context related to the buggy line. We provide details on our ablation results to aid future researchers in understanding which variables are most likely to improve their own models.

Due to randomness in learning, for each parameter, we run each configuration multiple times and report the mean and standard deviation for the model as recommended for assessment of random algorithms \cite{Arcuri2011}. As our goal is to select the best model for use in our Defects4J evaluation, we use the test set from CodRep4Full to select the best run of each model, hence we report the percentage decrease of the best run for a given model from the best result found with the golden model. Due to computational constraints, we only run each model 10 times; for the 18 configurations reported, almost 200GB of disk storage was used and 400 machine-hours. When using \approach to learn new datasets, we would recommend a similar approach where a validation set is used to select the best performing model after multitple training runs.

First, we consider the very coarse grain features.
Table~\ref{tab:buildup} shows the performance of four models, starting from
a simplistic seq-to-seq model that only takes a single buggy line $b_l$ as input when learning to produce the fixed line $f_l$. Then we show beam search, copy, and the use of the \textit{abstract buggy context} improving the model performance. These results confirm our answer to RQ2 that the copy mechanism is essential to the performance of the system.

Second, Table~\ref{tab:goldAblation} shows the results of our 'Golden model' against the results of single specific, targeted changes made to the model. 
Ablation ID 1 shows that our 10K training limit is sufficient given our training data. 
ID 2 shows that a vocabulary smaller than 1K tokens performs worse - likely due to a loss of learned tokens that can be used even if an instance of the token is not in the \textit{abstract buggy context}. 
ID 3 shows that a vocabulary larger than 1K tokens performs worse - perhaps due to the additional tokens having insufficient training examples for learning a proper embedding. To further understand the effect of vocabulary size, we analyzed the raw output of our model before the patch preparation step. For the golden model (vocab=1000), 38\% of the generated patches on CodRep4 have <unk> tokens and would be discarded; with ID 2 (700) it is 43\%, and with ID 3 (1400) it is 37\%. Hence, although a larger vocabulary had fewer raw <unk> tokens, the 1000 token vocabulary was able to produce better optimized models.

ID 4 is about pretraining; in order to provide more opportunities to learn a quality embedding, we created unsupervised pretraining data for the encoder/decoder. Using this unsupervised data did not improve the model, it worsened it.

ID 5 a and b show the value of combining the CodRep and Bugs2Fix data sets to improve the generalization of the model. ID 6 demonstrates the effect of removing the bridge between the encoder and decoder, which improved the mean for the model but tightened the standard deviation and hence produced a lower best result that the golden model. This is perhaps due to the bridge layer allowing for more variation in the encoder hidden state embedding and decoder hidden state embedding.

IDs 7 through 10 demonstrate that our LSTM network is sized correctly; presumably a smaller network cannot generalize on the model data well enough whereas a larger network has too many degrees of freedom. Our speculation is that a 2 layer encoder/decoder network allows the layer connected directly to the token embedding to 'focus' the weight matrix on input syntax while the layer connected to the attention/copy mechanism 'focuses' on output generation. 
ID 11 shows the loss in accuracy when \textit{abstract buggy context} is reduced to just the buggy line. 

ID 12 shows that truncation is necessary otherwise an out-of-memory error crashes the system, due to too many time steps being stored in memory per token in the sequence. ID 13 shows that if we truncated to 4,000 tokens then the system passes, but the increased context size (4,000 vs the golden model 1,000) did not improve accuracy of the model. 
ID 14 shows that using a 500 token limit for \textit{abstract buggy context} hurts accuracy presumably because there are less opportunities for token copy. We also speculate that a possible advantage of 1K truncation instead of 500 could be that 1K provides a type of unsupervised learning for the encoder hidden states, the global attention, and the copy mechanism.

ID 15 removes the <START\_BUG> and <END\_BUG> tokens from the \textit{abstract buggy context} input. The target output is still the correct single-line patch. Without these labels, \approach must learn line break positions and learn a type of fault localization in order to create a valid patch. Because \textit{abstact buggy context} does not include test coverage data or other information useful for fault localization, there is a significant accuracy loss for this ID, but the network was still able to create 356 correct patches. 

Our primary use case modeled in this paper is to use our golden model for \approach on projects for which it was not trained. This allows for a simpler use model than retraining the model periodically on an ongoing project. ID 16 explores the use case where \approach is trained with samples from the same projects that the buggy test cases come from. CodRep4 is added to the training set data and then 4,711 random samples are removed for testing (these samples may be from CodRep or Bugs2Fix project files). When the training data includes bugs from the same projects as the test data, we see a 12\% improvement in the best model. This use model is viable, but it does require more complete integration of \approach into a project regression system.

\begin{table}
\begin{center}
\begin{tabular}{ |p{5.5cm}|p{1cm}|p{0.9cm}| } 
 \hline
 Model description & CR4Full & ratio \\ 
 \hline
 50K vocab, no copy, beam size 1, no context & 55 & baseline \\ 
  50K vocab, no copy, beam size 50, no context & 206 & 3.7x \\
    1K vocab, copy, beam size 50, no context & 826 & 15.0x \\ 
 \hline
  Golden Model (with \textit{abstract buggy context}) & 950/4711 & 17.3x \\
  \hline
\end{tabular}
\caption{Performance impact of the key features of beam size, copy, and context.}
\label{tab:buildup}
\end{center}
\end{table}

\begin{table}
\begin{center}
\begin{tabular}{ |p{0.2cm}|p{3.9cm}|p{0.6cm}|p{0.3cm}|p{0.4cm}|p{0.65cm}| } 
 \hline
 ID & Model description & mean & SD &  max & chng \\ 
 \hline
 0 & Golden Model & 859 & 61 & 950 & --- \\ 
 \hline
 1 & more training iterations (20K vs 10K) & 832 & 78 & 901 & -5\% \\
 2 & smaller token vocabulary (700 vs 1000) & 824 & 70 & 886 & -7\% \\
 3 & larger token vocabulary (1400 vs 1000) & 868 & 32 & 907 & -5\% \\
 4 & with unsupervised pretraining & 821 & 65 & 922 & -3\% \\
 5a & less training data (CR vs CR+Bugs2Fix) & 742 & 47 & 810 & -15\% \\
 5b & less training data (Bugs2Fix vs CR+Bugs2Fix) & 748 & 24 & 785 & -17\% \\
 6 & no bridge layer from encoder to decoder & 887 & 34 & 942 & -1\% \\
 7 & fewer LSTM layers on enc/dec (1 vs 2) & 281 & 203 & 513 & -46\% \\
 8 & more LSTM layers on enc/dec (3 vs 2) & 833 & 49 & 914 & -4\% \\
 9 & fewer LSTMs per layer (128 vs 256) & 848 & 40 & 888 & -11\% \\
 10 & more LSTMs per layer (512 vs 256) & 812 & 89 & 907 & -5\% \\
 11 & without context (input only buggy line) & 738 & 63 & 826 & -13\% \\
 12 & no truncation of \textit{abstract buggy context} & crash & & & \\
 13 & truncate to larger context (4K vs 1K) & 848 & 79 & 950 & -0\% \\
 14 & truncate to smaller context (500 vs 1K) & 826 & 54 & 890 & -6\% \\
 15 & remove START\_BUG \& END\_BUG & 331 & 33 & 412 & -57\% \\
 16 & Intraproject training (4,711 testcases from CR+Bugs2Fix) & 984 & 47 & 1068 & +12\% \\

 \hline
\end{tabular}

\caption{Results with selected configurations in the parameter neighborhood of the golden model. For ID 0 through 15, results are total exact matches when model is tested on 4,711 testcases from CR4Full. ID16 results selected 4,711 testcases after merging CR1,2,3,4, and 5 with Bugs2Fix.}
\label{tab:goldAblation}
\end{center}
\end{table}

\section{Related Work}
\label{sec:related}
The work presented here is on built on top of two big and active research fields: program repair and machine learning on code. We refer to recent surveys for getting a good overview on them: \cite{Monperrus2015} for program repair and Allamanis \etal's \cite{allamanis2018survey} for the latter.
In the following, we focus on those works that are about learning and automatic repair.

sk\_p is a program repair technique for syntactic and semantic errors in student programs submitted to MOOCs \cite{Pu:2016}. First, it uses the previous and next statement to predict the statement in the middle, \ie to replace the current statement. The probability of a patch is the product of the probabilities for all chosen statements. As we do, sk\_p uses beam search to produce the top n predictions. 

Another paper on MOOCs \cite{Bhatia2018} repairs student submissions in Python by combining learning and sketch-based synthesis. 
The approach by Wang \etal \cite{wang2017dynamic} considers MOOC but the technique itself is completely different: \cite{wang2017dynamic} does deep learning on program traces in order to predict the kind of bug affecting a student submission.
The main differences between those works and ours are that 1) we consider a larger context (the buggy class) and 
2) we consider real programs for training and testing that are bigger and more complex than student's submissions.
Shin \etal \cite{shin2018towards} consider simple programs in the educational programming language Karel. As \approach, their system predicts to delete, insert or replace tokens. 
Henkel \etal \cite{henkel2018code} compute an embedding for symbolic traces and perform a pilot experiment for fixing error-handling code, which is very different from concrete bug fixing as we do here.

DeepFix is a program repair tool for fixing compiler errors in introductory programming courses \cite{gupta2017deepfix}. The input is the whole program, (100 to 400 tokens long for their data), and the output is a single line fix. The vocabulary size is set  to 129, which was enough to map every distinct token type to a unique word in the vocabulary. 
TRACER is another program repair tool for fixing compiler errors which outperforms DeepFix in terms of success rate \cite{ahmed2018compilation}. 
Santos \etal's \cite{santos2018syntax} further refines the idea and evaluates it with an even larger dataset.
The focus of those three works and ours is very different, they focus on compiler errors, we focus on logical bugs. For compiler errors, one does not need to consider the whole vocabulary, but only token types. On the contrary, we have to address this problem and we do so by using the copy mechanism.

DeepRepair \cite{DBLP:journals/corr/WhiteTMMP17} is an early attempt to integrate machine learning in a program repair loop. 
DeepRepair leverages learned code similarities, captured with  recursive autoencoders \cite{White:2016:DLC:2970276.2970326}, to select repair ingredients from code fragments that are similar to the buggy code. Our usage of learning is different, DeepRepair uses machine learning to select interesting code, \approach uses machine learning to generate the actual patch.

Tufano \etal investigated the feasibility of using neural machine translation for learning bug-fixing patches via NMT \cite{tufano2018arxiv}. The authors first perform a source code abstraction process that relies on a combination of Lexer+Parser which replaces identifiers and literals in the code.
The goal of this abstraction is it reduce the vocabulary while keeping the most frequent identifiers/literals.
In their work the authors analyzed small methods (no longer than 50 tokens) and medium methods (no longer than 100 tokens) and observed a drop in performance for longer methods. Since their approach takes a buggy method as input and generates the entire fixed method as output, the maximum method length Tufano \etal considered is only 100 tokens. Their work addressed the vocabulary problem by renaming rare identifiers through a custom abstraction process.
\approach is different in the following ways.
First, we consider the entire context of the buggy class, rather than only the buggy method, in order for the model to access more tokens when predicting the fix. 
Second, our abstraction process uniquely utilizes the copy mechanism (which they do not), which allows \approach to utilize a larger set of tokens when generating the fix and to include information about the context within the \textit{abstract buggy context} in which a token is used. 
Beyond those two major qualitative differences, a quantitative one is that they only consider small methods, no longer than 100 tokens, while we have no such restriction; \approach can potentially generate a one-line patch within a method of any size.

Parallel work by Hata \etal \cite{Hata2018} discusses a similar network architecture, also applied to one-line diffs.
The major differences between \cite{Hata2018} and our work are the following:
First, they do project-specific training, which means that their approach is only evaluated on testing data coming from the same project. On the contrary, we do global training and we show that \approach captures repair operators applicable to any project. Our qualitative case studies are unique with that respect.
Second, they only look at wellformedness of the output, while we also compile and execute the predicted patch. Our work is an end-to-end test-suite based repair approach. Third, their input is limited to the precise buggy code to replace, while \approach uses \textit{abstract buggy context}, which allows for a broader set of tokens for the copy mechanism to select from.

\section{Conclusion}

In this paper, we have presented a novel approach to program repair, called \approach, based on sequence-to-sequence learning. Our approach uniquely combines an encoder/decoder architecture with the copy mechanism to overcome the problem of large vocabulary in source code. On a testing dataset of 4,711 tasks taken from projects which were not in the training set, \approach is able to successfully predict 950 changes. On Defects4J one-line bugs, \approach produces 61 plausible, test-suite adequate patches. To our knowledge, our paper is the first ever to show the effectiveness of the copy mechanism for program repair, which provides a mechanism
to alleviate the unlimited vocabulary problem.

This work opens promising research directions.
First, we aim to improve and adapt \approach with the goal of addressing multi-line patches. We believe there are different ways we can tackle this: (i) for fixes modifying contiguous lines of code (\ie hunk) we can extend \approach to learn to generate multiple lines of code as output, with the special tokens (\ie \texttt{<START\_BUG>} and \texttt{<END\_BUG>}) surrounding the entire hunk; (ii) for fixes modifying multiple lines in different locations, we could envision \approach generating a finite set of combinations of the program containing a predicted fixed line for each of the suspicious locations.
Second, there is some preliminary work on tree-to-tree transformation learning \cite{abs-1810-00314}, which conceptually is very appropriate for code viewed as parse trees. Such techniques may augment or supersede sequence-to-sequence approaches.
Finally, the originality of our context abstraction is to capture class-level, long range dependencies: we will study whether such a network architecture is able to capture dependencies beyond that, at the package or application level.

%
%
%
%
%
%

%
%
%
%
%
%

%
%
%
%
%
%
%

%

\printbibliography

\begin{IEEEbiography}[{\includegraphics[width=1in,height=1.25in,clip,keepaspectratio]{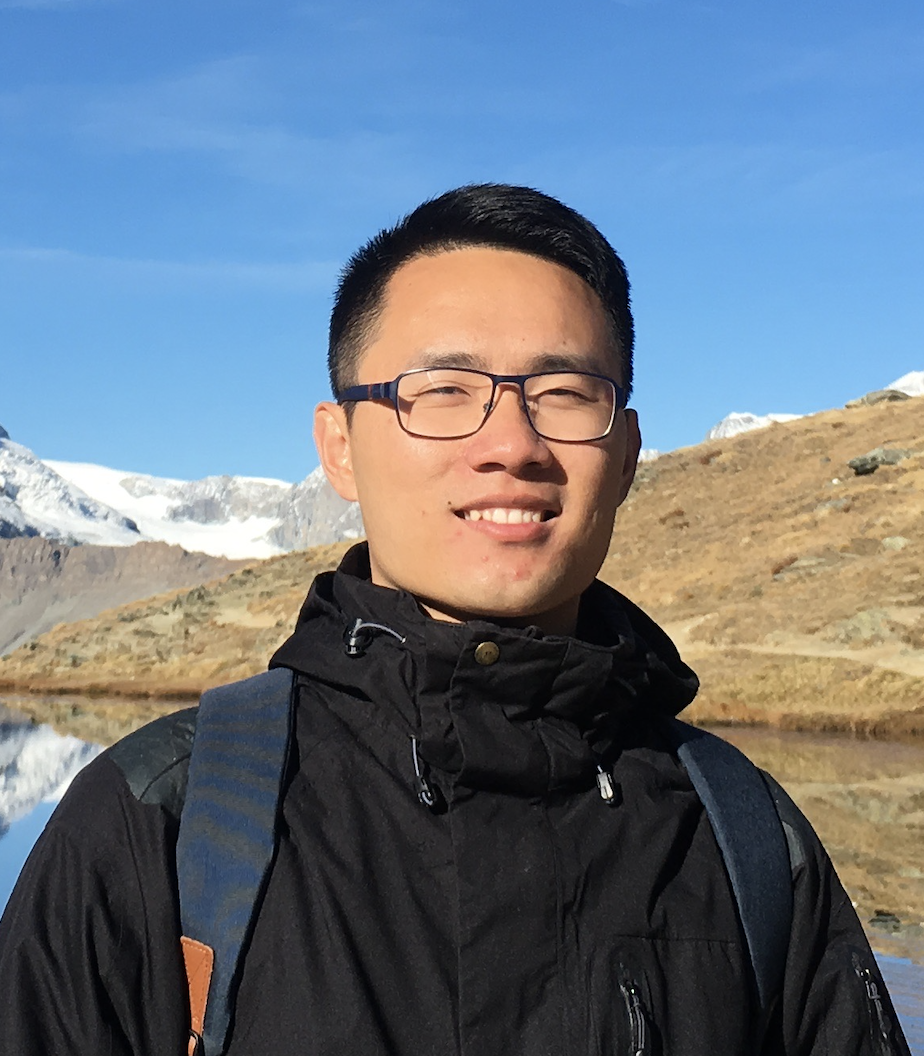}}]{Zimin Chen} is currently a PhD student at KTH Royal Institute of Technology. He also received the BS and MS degree in computer science from KTH. His research interest lies in the intersection between machine learning and software engineering, especially between automatic program repair and machine learning.

\end{IEEEbiography}

\vspace{-1cm}
\begin{IEEEbiography}[{\includegraphics[width=1in,height=1.25in,clip,keepaspectratio]{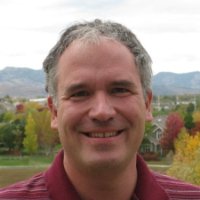}}]{Steve Kommrusch} is currently a PhD candidate focused on machine learning at Colorado State University. He received his BS in computer engineering from University of Illinois in 1987 and his MS in EECS from MIT in 1989. From 1989 through 2017, he worked in industry at Hewlett-Packard, National Semiconductor, and Advanced Micro Devices. Steve holds over 30 patents in the fields of computer graphics algorithms, silicon simulation and debug techniques, and silicon performance and power management. His research interests include Program Equivalence, Program Repair, and Constructivist AI using machine learning. More information available at: https://www.cs.colostate.edu/~steveko/.
\end{IEEEbiography}

\vspace{-1cm}
\begin{IEEEbiography} [{\includegraphics[width=1in,height=1.25in,clip,keepaspectratio]{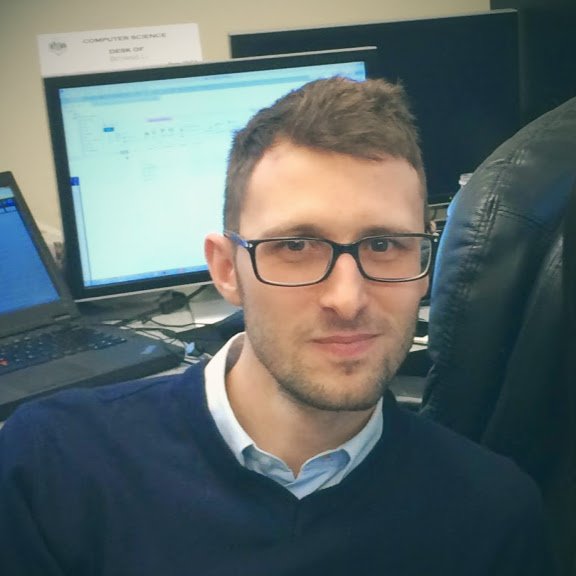}}] {Michele Tufano}
received his Ph.D. in Computer Science from The College of William \& Mary in May 2019. He received a B.S. in Computer Science from the University of Salerno in 2012 and his M.S. in Computer Science from the University of Salerno in 2014. His research interests include Deep Learning applied to Software Engineering, Automated Program Repair, Software Evolution and Maintenance, Mining Software Repositories, and Android Testing.  He received an ACM SIGSOFT Distinguished Paper Award at ICSE 2015. More information available at: \url{https://tufanomichele.com/}.
\end{IEEEbiography}

\vspace{-1cm}
\begin{IEEEbiography}[{\includegraphics[width=1in,height=1.25in,clip,keepaspectratio]{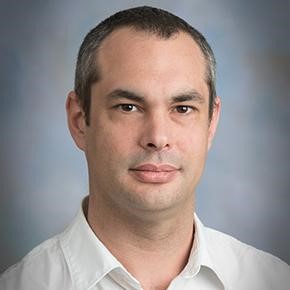}}]
{Louis-No{\"e}l Pouchet}
Dr. Louis-Noel Pouchet is an Associate Professor of Computer Science at Colorado State University, with a joint appointment in the Electrical and Computer Engineering department. He is working on pattern-specific languages and compilers for scientific computing, and has designed numerous approaches using optimizing compilation to effectively map applications to CPUs, GPUs, FPGAs and System-on-Chips. His work spans a variety of domains including compiler optimization design especially in the polyhedral compilation framework, high-level synthesis for FPGAs and SoCs, and distributed computing. Previously he has been a Visiting Assistant Professor (2012-2014) at the University of California Los Angeles, where he was a member of the NSF Center for Domain-Specific Computing, working on both software and hardware customization, and a Research Assistant Professor at the Ohio State University till 2016. He is the author of the PolyOpt and PoCC compilers, and of the PolyBench benchmarking suite.
\end{IEEEbiography}

\vspace{-1cm}
\begin{IEEEbiography}
[{\includegraphics[width=1in,height=1.25in,clip,keepaspectratio]{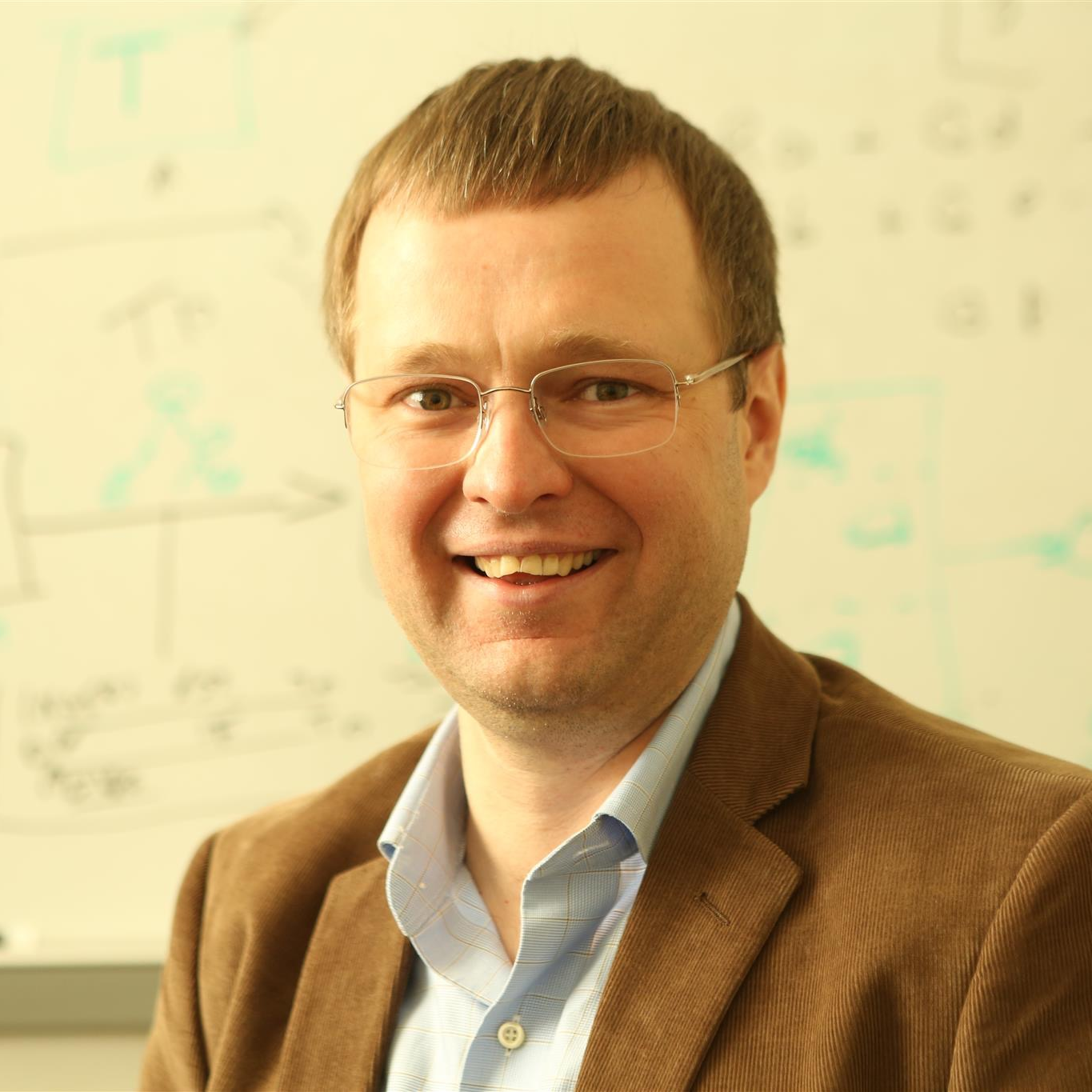}}]
{Denys Poshyvanyk}
 is the Class of 1953 Term Distinguished Associate
Professor of Computer Science at the College of William and Mary in
Virginia. He received the MS and MA degrees in Computer Science from
the National University of Kyiv-Mohyla Academy, Ukraine, and Wayne
State University in 2003 and 2006, respectively.

He received the PhD degree in Computer Science from Wayne State
University in 2008. He served as a program co-chair for MobileSoft'19,
ICSME'16, ICPC'13, WCRE'12 and WCRE'11. He currently serves on the
editorial board of IEEE Transactions on Software Engineering (TSE),
Empirical Software Engineering Journal (EMSE, Springer) and Journal of
Software: Evolution and Process (JSEP, Wiley). His research interests
include software engineering, software maintenance and evolution,
program comprehension, reverse engineering, software repository
mining, source code analysis and metrics. His research papers received
several Best Paper Awards at ICPC'06, ICPC'07, ICSM'10, SCAM'10,
ICSM'13 and ACM SIGSOFT Distinguished Paper Awards at ASE'13, ICSE'15,
ESEC/FSE'15, ICPC'16, ASE'17 and FSE'19. He also received the Most
Influential Paper Awards at ICSME'16 and ICPC'17. He is a recipient of
the NSF CAREER award (2013).  He is a member of the IEEE and ACM. More
information available at: \url{http://www.cs.wm.edu/~denys/}.

\end{IEEEbiography}

\vspace{-1cm}
\begin{IEEEbiography}
[{\includegraphics[width=1in,height=1.25in,clip,keepaspectratio]{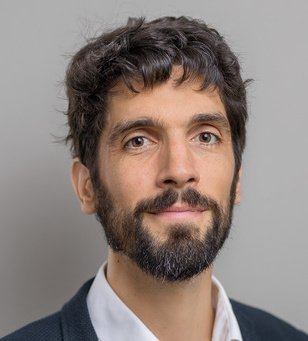}}]
{Martin Monperrus}
is Professor of Software Technology at KTH Royal Institute of Technology. He was previously associate professor at the University of Lille and adjunct researcher at Inria. He received a Ph.D. from the University of Rennes, and a Master's degree from the Compiègne University of Technology. His research lies in the field of software engineering with a current focus on automatic program repair, program hardening and chaos engineering.
\end{IEEEbiography}

\end{document}